\documentclass[aps,pre,twocolumn,reprint,groupedaddress]{revtex4}
\usepackage{graphicx}
\usepackage{amsmath}
\usepackage{mathtools}
\usepackage{bbold}
\usepackage[usenames, dvipsnames]{color}

\usepackage{hyperref}

\usepackage{xcolor}

\allowdisplaybreaks

\begin{document}

\title{Calculation of critical exponents on fractal lattice Ising model by higher-order tensor renormalization group method}

\author{Jozef \textsc{Genzor}\footnote{~jozef.genzor@gmail.com}}
\affiliation{Physics Division, National Center for Theoretical Sciences, National Taiwan University, Taipei 10617, Taiwan}

\date{\today}

\begin{abstract}
The critical behavior of the Ising model on a fractal lattice, which has the Hausdorff dimension $\log_{4} 12 \approx 1.792$, is investigated using a modified higher-order tensor renormalization group algorithm supplemented with automatic differentiation to compute relevant derivatives efficiently and accurately.
The complete set of critical exponents characteristic of a second-order phase transition was obtained. 
Correlations near the critical temperature were analyzed through two impurity tensors inserted into the system, which allowed us to obtain the correlation lengths and calculate the critical exponent $\nu$. 
The critical exponent $\alpha$ was found to be negative, consistent with the observation that the specific heat does not diverge at the critical temperature. 
The extracted exponents satisfy the known relations given by various scaling assumptions within reasonable accuracy. 
Perhaps most interestingly, the hyperscaling relation, which contains the spatial dimension, is satisfied very well, assuming the Hausdorff dimension takes the place of the spatial dimension. 
Moreover, using automatic differentiation, we have extracted four critical exponents ($\alpha$, $\beta$, $\gamma$, and $\delta$) globally by differentiating the free energy. 
Surprisingly, the global exponents differ from those obtained locally by the technique of the impurity tensors; however, the scaling relations remain satisfied even in the case of the global exponents.
\end{abstract}

\maketitle

\section{Introduction}

The phase transition and critical phenomena are prominent topics in condensed matter physics~\cite{Domb_Green}.
The scaling behavior of physical quantities, such as the magnetic susceptibility and specific heat, is characterized by the critical exponents when approaching the critical temperature~\cite{Baxter}.
Various scaling assumptions give the relations between the critical exponents. 
One of such relations derived from the \textit{hyperscaling hypothesis}, which is expected to be valid for $d \leq 4$, involves the system dimension $d$.
An intriguing question is the validity of the hyperscaling hypothesis in the case of non-integer dimensional systems such as fractals, where critical phenomena remain understudied. 

\begin{figure}[tb]
\includegraphics[width=0.48\textwidth,clip]{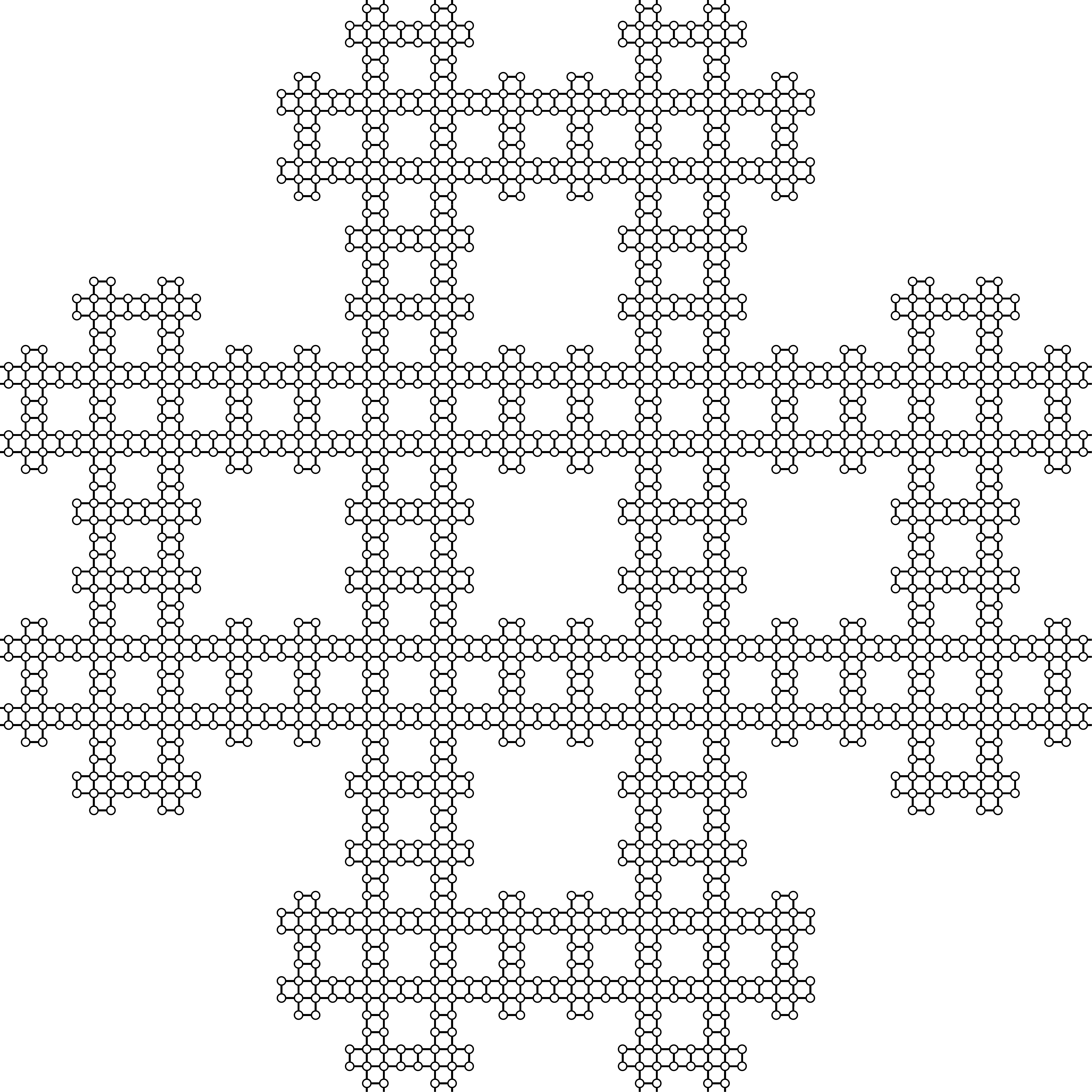}
\caption{
The layout of the fractal lattice after three extension steps, $n=3$. 
Tiny circles represent the two-state Ising spins. 
The horizontal and vertical lines represent the spin-spin interactions.
The number of sites grows as $12^n$ with the number of extension steps $n$, whereas the number of outgoing bonds grows as $2^{n+2}$.
}
\label{fig:Fig_1}
\end{figure}

To some extent, the hyperscaling relation expressed in terms of the ratios of the critical exponents ${\beta} / {\nu}$ and ${\gamma} / {\nu}$ has already been considered in the literature
\begin{equation} \label{d_eff}
d_{\rm eff}^{~} = 2\dfrac{\beta}{\nu} + \dfrac{\gamma}{\nu} \, , 
\end{equation}
where $d_{\rm eff}$ is the effective dimension that controls hyperscaling. 
Nevertheless, the question of whether the effective dimension $d_{\rm eff}$ is the same as the Hausdorff dimension $d_{\rm H}$ remains open for debate. 
The validity of the hyperscaling relation was mostly tested in the case of Sierpi\'{n}ski carpets. 
The Ising model on Sierpinski carpets SC$(3,1)$ of Hausdorff dimension $d_{\rm H} = \ln 8/ \ln 3 \approx 1.8927$ and SC$(4,2)$ of Hausdorff dimension $d_{\rm H} = \ln 12 / \ln 4 \approx 1.7924$ was studied using Monte Carlo in conjunction with the finite-size scaling method in Ref.~\cite{Carmona}. 
The existence of an order-disorder transition at finite temperature was clearly shown in both cases, and the critical exponents, including their errors, were estimated.
In this case, the hyperscaling relation holds if one assumes that the effective dimension is the Hausdorff dimension. 
In case of SC$(3,1)$, the exponent $\alpha$, for which the hyperscaling relation reads $\nu d_{\rm H} = 2 - \alpha$, was found to be negative. 
Consistent with the previous conclusions is a newer Monte Carlo study with finite-size scaling analysis in Ref.~\cite{Monceau} where the authors studied four different Sierpinski carpets with the Hausdorff dimension $d_{\rm H}$ between $1.9746$ and $1.7227$, namely SC$(5,1)$, SC$(3,1)$, SC$(4,2)$, and SC$(5,3)$. 
In the case of SC(3,1), the authors found $d_{\rm eff}$ to be only slightly smaller than $d_{\rm H}$. 
In Ref.~\cite{Bab1}, the short-time dynamic evolution of an Ising model on Sierpinski carpet SC$(3,1)$ was studied using the Monte Carlo method.
The authors concluded that the effective dimension for the second order phase transition is noticeably smaller than the Hausdorff dimension $d_{\rm eff} \sim 1.77 < d_{\rm H}$. 
Another short-time critical dynamic scaling study in the case of various infinitely ramified fractals with Hausdorff dimension within the interval $1.67 \leq d_{\rm H} \leq 1.98$ can be found in Ref.~\cite{Bab2}. 
Their results are consistent with the convergence of the lower-critical dimension toward $d=1$ for fractal substrates and suggest that the Hausdorff dimension may differ from the effective dimension.
The values for the different sets of fractals depart from $d_{\rm eff} = d_{\rm H}$ for $d_{\rm H} \leq 1.85$.
However, due to large error bars, the authors cannot state a definitive conclusion on the actual dependence of $d_{\rm eff}$ on $d_{\rm H}$.
The Ising model on a fractal lattice with Hausdorff dimension $d_{\rm H} = \ln 12 / \ln 4 \approx 1.792$ (which is different from SC$(4,2)$ of the same $d_{\rm H}$) depicted in Fig.~\ref{fig:Fig_1} was already probed by two different adaptations of the Higher-Order Tensor Renormalization Group (HOTRG, introduced in Ref.~\cite{HOTRG}):
(1) genuine fractal representation (with no structure filling the gaps)~\cite{2dising, APS}, and 
(2) $J_1$-$J_2$ (``tunable'') fractal constructed on a square-lattice frame with two types of couplings, $J_1$ and $J_2$~\cite{j1j2}. 
Geometrically, these two methods constitute the same fractal when $(J_1, J_2) = (1, 0)$ albeit represented differently. 
Two critical exponents, $\beta$ and $\delta$, were extracted in both cases using the technique of local impurity tensors. 
There is a slight discrepancy between the values of the exponents between the two methods, which can be attributed to the difference in the details of the calculation of the local magnetization rather than the model representation. 
The magnetization in (1) is calculated on a single site located far from the system's external boundary, whereas in (2), a partial average over central sites is employed. 
This discrepancy is interesting since it indicates that there is a positional dependency, at least in the case of local magnetization. 
Another interesting finding is that the specific heat does not exhibit singular behavior around its maximum; however, a sharp peak was observed in a numerical derivative of the specific heat at the critical temperature. 
The question about the value of the critical exponent $\alpha$ associated with the specific heat has remained open until now. 
%

The position dependence of local thermodynamic functions was studied in Ref.~\cite{carpet}, where HOTRG was adapted to the classical Ising model on SC$(3,1)$. 
The critical temperature $T_{\rm c}^{~}$ was found to be positionally independent, whereas the (local) critical exponent $\beta$ was found to vary by two orders of magnitude depending on lattice location. 
Let us mention that the Monte Carlo studies achieve only a relatively modest maximal value of the segmentation steps $k \leq 8$.
In contrast, in the case of the Higher Order Tensor Renormalization Group (HOTRG) method used in the study of SC$(3,1)$ in Ref.~\cite{carpet}, numerical convergence of the physical observables is achieved at $k \sim 35$ iterative extensions (generations) of the system. 
%

In Ref.~\cite{gasket}, the quantum phase transition of the transverse-field Ising model on the Sierpi\'{n}ski fractal with the Hausdorff dimension $\log_2^{~} 3 \approx 1.585$ was studied by a modified HOTRG method. 
Ground-state energy and order parameter were calculated and analyzed.   
The system was found to exhibit a second-order phase transition.
From the order parameter, the critical exponents $\beta$ and $\delta$ were estimated. 
%

Recently it was shown that the higher-order derivatives for the tensor network algorithms could be calculated accurately and efficiently using the technique of automatic differentiation~\cite{ad1, ad2}, which emerged from deep learning. 
Automatic differentiation is based on the concept of the computation graph, which is a directed acyclic graph composed of elementary computation steps.
This technology propagates the gradients through the whole computation process with machine precision. 
In the case of the tensor network algorithms, an essential technical ingredient is to implement numerically stable differentiation through linear algebra operations such as the Singular Value Decomposition (SVD). 
Applying the automatic differentiation on our tensor network fractal, we can now calculate specific heat very accurately as the first derivative of the bond energy with respect to temperature or as a second derivative of the free energy with respect to temperature without introducing numerical errors due to the finite step as in the case of numerical derivatives. 
With such an accurate method to obtain the specific heat, it should be possible to extract the associated critical exponent $\alpha$ finally. 
Similarly, the magnetic susceptibility can now be calculated as a first derivative of the spontaneous magnetization with respect to the external field or as a second derivative of the free energy with respect to the external field. 
Having calculated the magnetic susceptibility, one can extract the critical exponent $\gamma$.
Finally, let us emphasize, that differentiation of the free energy would yield global thermodynamic quantities, which were not calculated before. 

%
A question of high interest is to numerically estimate the critical exponent $\nu$, which appears in the hyperscaling relation together with the spatial dimension $d$. 
The critical exponent $\nu$ can be extracted from the correlation length, which can be obtained from the correlation function. 
The method for calculation of the correlation function using the Tensor Renormalization Group (TRG) method was introduced in Ref.~\cite{TRG}.
This method was implemented and tested in the case of the square lattice Ising model in Ref.~\cite{Correlations}.
A similar approach is conceivable in the case of the HOTRG method; therefore, it can be used on the fractal lattice under study. 

In this study, we have extracted the remaining four critical exponents from our HOTRG calculations with the local impurity tensors by calculating the correlation function for obtaining the exponents $\nu$ and $\eta$ and by augmenting our computations with the automatic differentiation for $\alpha$ and $\gamma$.   
The critical exponent $\alpha$ was found to be negative ($\alpha \approx -0.87$), which is consistent with the observation that the specific heat does not diverge at the critical temperature. 
The exponents we extracted satisfy the known relations given by various scaling assumptions with reasonable accuracy. 
Perhaps most interestingly, the hyperscaling relation, which contains the spatial dimension, is satisfied very well, assuming the Hausdorff dimension takes the place of the spatial dimension. 
Moreover, using automatic differentiation, we have extracted four critical exponents ($\alpha$, $\beta$, $\gamma$, and $\delta$) globally by differentiating the free energy. 
Surprisingly, the global exponents are very different from those obtained locally by the technique of the impurity tensors (for example, the global exponent $\beta$ is more than five times larger than the local $\beta$); however, the scaling relations remain satisfied even in the case of the global exponents.

\section{Model representation}
We consider the nearest-neighbors fractal-lattice Ising model with the Hamiltonian
\begin{equation}
H = - J \displaystyle\sum_{\left<i j\right>}\sigma_{i}^{~} \sigma_{j}^{~} - h \displaystyle\sum_{i}^{~} \sigma_{i}^{~} \, ,
\end{equation}
where $J>0$ is the ferromagnetic coupling, and $h$ is the uniform magnetic field. 
At each site $i$, the Ising variable $\sigma_i^{~}$ takes only two values, $+1$ or $-1$. 
For brevity, we set $J=1$ and $h=0$ in the following. 
The partition function of the Ising model defined on the fractal lattice can be expressed in terms of tensor network states defined by four types of local tensors represented by $T$, $X$, $Y$, and $Q$,
%
%
\begin{eqnarray} 
\label{t0_def}
T_{x_i^{~} x'_i y_i^{~} y'_i} &=& \raisebox{-1.6em}{\includegraphics[height=3.5em]{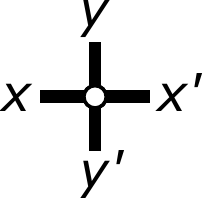}} = \displaystyle\sum_{\xi}^{~} W_{\xi x_i^{~}} W_{\xi x'_i} W_{\xi y_i^{~}} W_{\xi y'_i} \, ,  \\[-10pt]
\label{x0_def}
X_{x_i^{~} x'_i y_i^{~}} &=& \raisebox{-1.6em}{\includegraphics[height=3.5em]{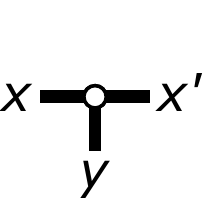}} = \displaystyle\sum_{\xi}^{~} W_{\xi x_i^{~}} W_{\xi x'_i} W_{\xi y_i^{~}} \, ,  \\[3pt]
\label{y0_def}
Y_{y_i^{~} y'_i x_i^{~}} &=& \raisebox{-1.6em}{\includegraphics[height=3.5em]{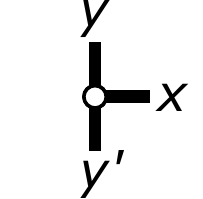}} = \displaystyle\sum_{\xi}^{~} W_{\xi y_i^{~}} W_{\xi y'_i} W_{\xi x_i^{~}} \, ,  \\[-10pt]
\label{q0_def}
Q_{x_i^{~} y_i^{~}} &=& \raisebox{-1.6em}{\includegraphics[height=3.5em]{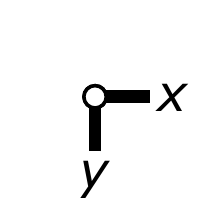}} = \displaystyle\sum_{\xi}^{~} W_{\xi x_i^{~}} W_{\xi y_i^{~}} \, , 
\end{eqnarray}
where $W$ is a $2\times2$ matrix determined by the bond weight factorization. 
While the choice for $W$ is arbitrary to a certain degree, here we choose an asymmetric factorization
%
%
\begin{equation} \label{weight}
W = \left(\begin{array}{lr} 
\sqrt{\cosh 1 / T} & \sqrt{\sinh 1 / T}   \\
\sqrt{\cosh 1 / T} & -\sqrt{\sinh 1 / T} \end{array} \right) \, ,
\end{equation}
where $T$ is the temperature. 
Notice that when two local tensors are contracted via non-physical (auxiliary) index $x$, the bond weight ${\cal{W}}_{\rm B} \left(\sigma_{i}, \sigma_{j}\right) = \exp{\left(\sigma_{i} \sigma_{j} / T \right)}$ is correctly re-expressed
\begin{equation}
{\cal{W}}_{\rm B} \left(\sigma_{i}, \sigma_{j}\right) = \sum_{x=0}^{1} W_{\xi_i x} W_{\xi_j x} \, ,
\end{equation}
where the first matrix index $\xi_i = (1 - \sigma_i) / 2$ takes values of $0$ and $1$ when $\sigma_i = 1$ and $\sigma_i = -1$, respectively.

The coarse-graining renormalization procedure introduced in Ref.~\cite{2dising, APS} is used to calculate the partition function. 
We start counting the iteration steps from zero; therefore, we denote the initial tensors in Eqs.~\eqref{t0_def}~--~\eqref{q0_def} as $T^{(n=0)}_{~} = T_{~}^{~}$, $X^{(n=0)}_{~} = X_{~}^{~}$, $Y^{(n=0)}_{~} = Y_{~}^{~}$, and $Q^{(n=0)}_{~} = Q_{~}^{~}$. 
At each iterative step $n$, the new tensors $T^{(n+1)}_{~}$, $X^{(n+1)}$, $Y^{(n+1)}$, and $Q^{(n+1)}_{~}$ are created from the previous-iteration tensors $T^{(n)}_{~}$, $X^{(n)}_{~}$, $Y^{(n)}_{~}$, and $Q^{(n)}_{~}$, according to the following extension relations
\begin{eqnarray}
T_{\left(x_1^{~} x_2^{~}\right) \left(x^{\prime}_1 x^{\prime}_2\right) \left(y_1^{~} y_2^{~}\right) \left(y^{\prime}_1 y^{\prime}_2\right)}^{(n+1)} \, &=& \,  \raisebox{-3.5em}{\includegraphics[height=7.5em]{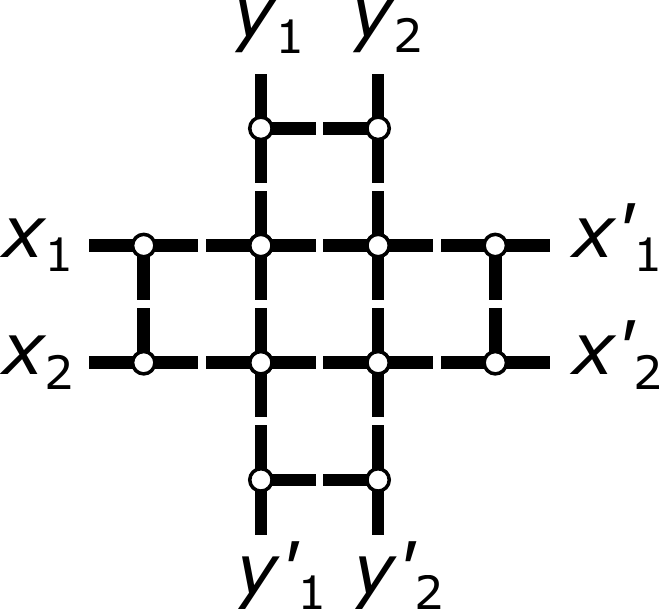}} \, , \label{T_ext} \\[-10pt]
X_{\left(x_1^{~} x_2^{~}\right) \left(x^{\prime}_1 x^{\prime}_2\right) \left(y_1^{~} y_2^{~}\right)}^{(n+1)} \, &=& \, \raisebox{-3.5em}{\includegraphics[height=7.5em]{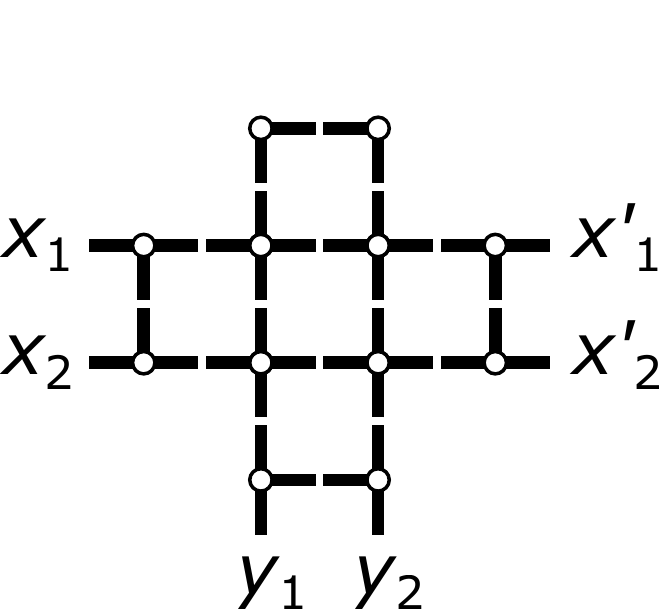}} \, , \\[3pt]
Y_{\left(y_1^{~} y_2^{~}\right) \left(y^{\prime}_1 y^{\prime}_2\right) \left(x_1^{~} x_2^{~}\right)}^{(n+1)} \, &=& \, \raisebox{-3.5em}{\includegraphics[height=7.5em]{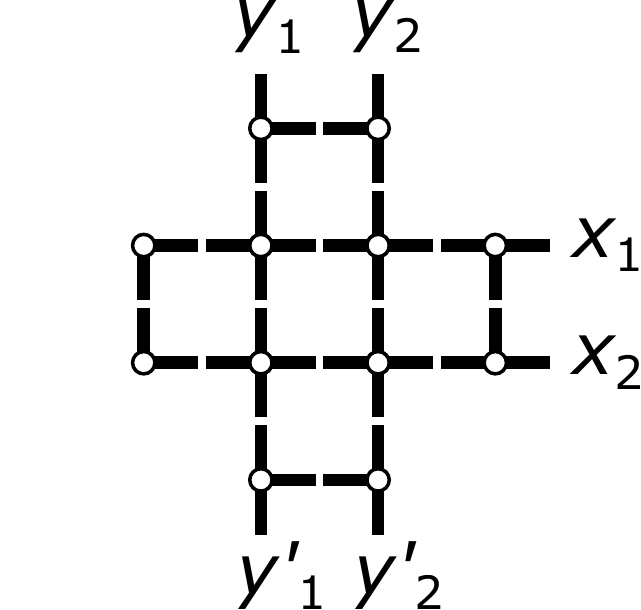}} \, , \\[-10pt]
Q_{\left(x_1^{~} x_2^{~}\right) \left(y_1^{~} y_2^{~}\right)}^{(n+1)} \, &=& \, \raisebox{-3.5em}{\includegraphics[height=7.5em]{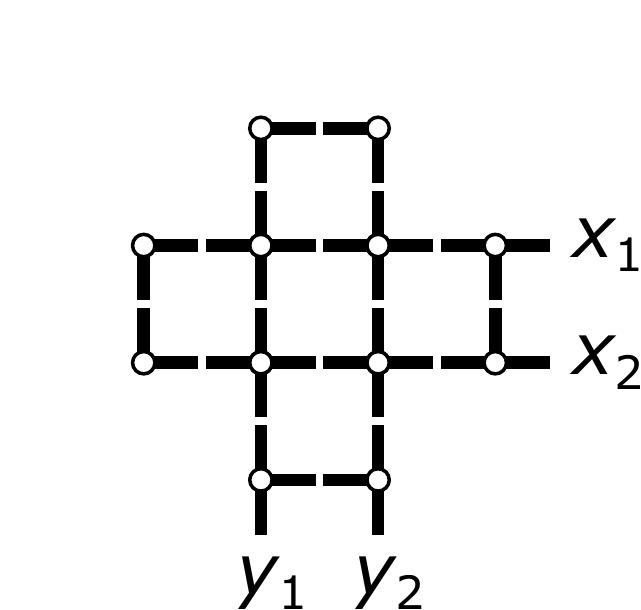}} \, .
\end{eqnarray}
The partition function $Z_{n}^{~} \left(T\right)$ of the system after $n$ extensions is evaluated as 
\begin{equation}
Z_{n}^{~} (T) = \displaystyle\sum_{ij}^{~} T^{(n)}_{iijj} \, , 
\end{equation}
where we impose the periodic boundary conditions. 

\subsection{Renormalization transformation}

At each iteration step, the bond dimension increases quadratically.
Therefore, a renormalization transformation is employed to limit the degrees of freedom kept at each tensor index. 
We update the local tensors by inserting horizontal (depicted by red dashed lines) and vertical projectors (depicted by blue dashed lines) into the extension relations
\begin{eqnarray}
T_{x_{~}^{~} x^{\prime}_{~} y_{~}^{~} y^{\prime}_{~}}^{n+1} \, &=& \,  \raisebox{-3.9em}{\includegraphics[height=8.2em]{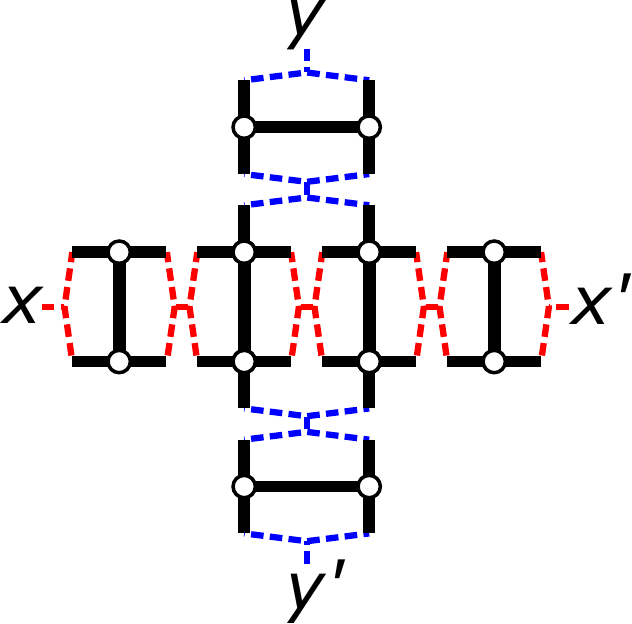}} \, , \\[-10pt]
X_{x_{~}^{~} x^{\prime}_{~} y_{~}^{~}}^{n+1} \, &=& \,  \raisebox{-3.9em}{\includegraphics[height=8.2em]{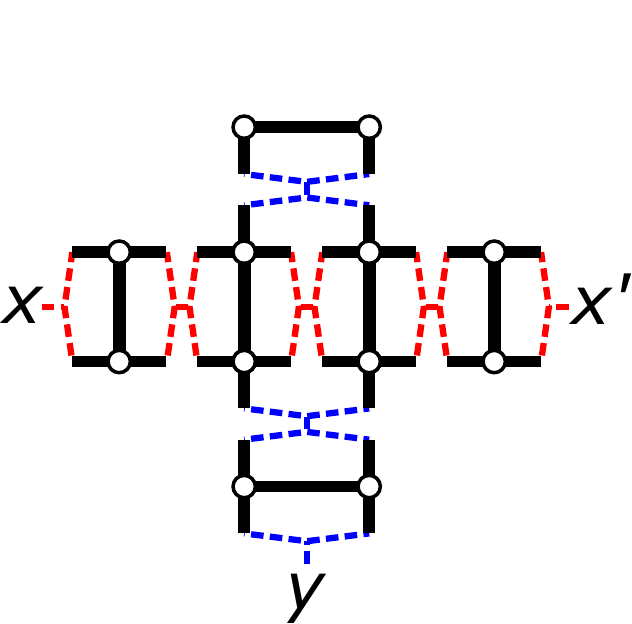}} \, , \\[3pt]
Y_{y_{~}^{~} y^{\prime}_{~} x_{~}^{~}}^{n+1} \, &=& \,  \raisebox{-3.9em}{\includegraphics[height=8.2em]{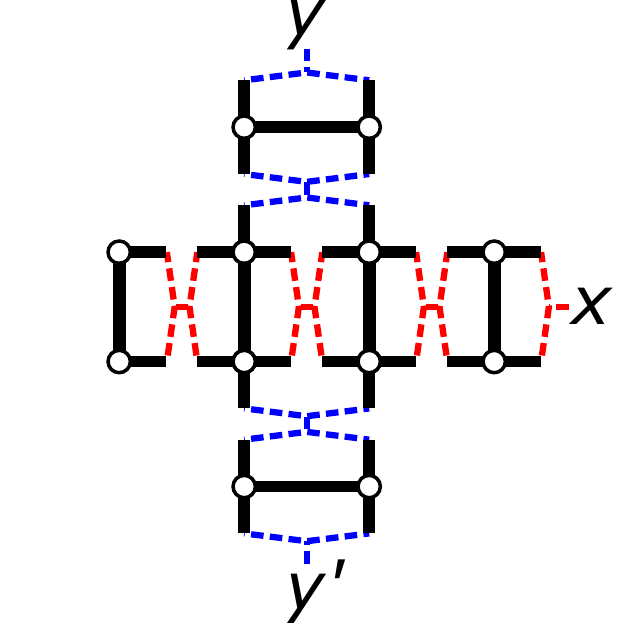}} \, , \\[-10pt]
Q_{x_{~}^{~} y_{~}^{~}}^{n+1} \, &=& \,  \raisebox{-3.9em}{\includegraphics[height=8.2em]{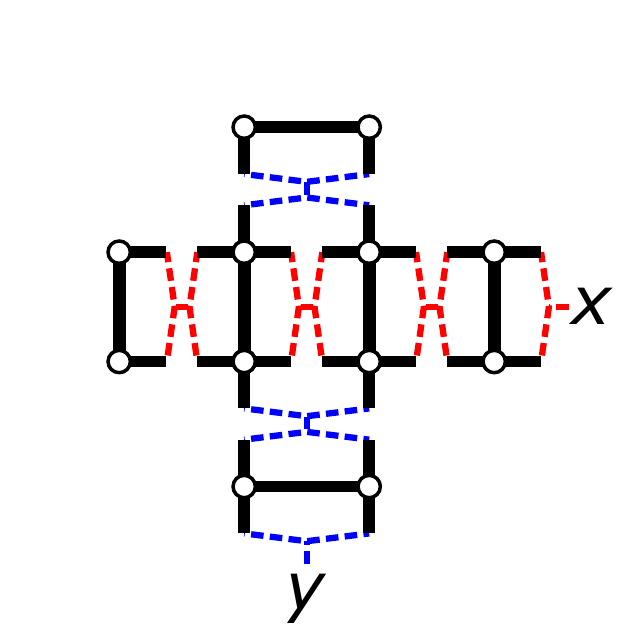}} \, .
\end{eqnarray}
The horizontal projector $U^{~}_{\rm X}$ is obtained from the Higher-Order Singular Value Decomposition (HOSVD)~\cite{hosvd} applied to 
\begin{equation}
M^{(n)}_{x x^{\prime} y y^{\prime}} = \sum_s^{~} T^{(n)}_{x_1^{~} x_1^{\prime} y s} T^{(n)}_{x_2^{~} x_2^{\prime} s y^{\prime}_{~}} \, ,
\end{equation}
where $x = x_1 \otimes x_2$ and $x^{\prime} = x^{\prime}_1 \otimes x^{\prime}_2$. 
Alternatively, we define the matrix unfolding $M^{\prime}_{x \, , \, \left(x^{\prime}_{~} y y^{\prime}_{~}\right)} = M^{(n)}_{x x^{\prime} y y^{\prime}}$ by regrouping the indices and then we perform an SVD
\begin{equation}
M^{\prime}_{~} = U^{~}_{\rm X} \omega_{\rm X} V^{\dag}_{\rm X} \, , 
\end{equation}
where $U^{~}_{\rm X}$ and $V^{\dag}_{\rm X}$ are unitary matrices, and $\omega_{\rm X}$ is a matrix with the singular values as its diagonal entries ordered in decreasing order by convention. 
The second index dimension of $U^{~}_{\rm X}$ is truncated down to the $D$, where $D$ is the dimension threshold of the truncated tensor dimension. 
Next, we will obtain the vertical projector $U^{~}_{\rm Y}$.
To do that, we first prepare a truncated tensor
\begin{equation}
\tilde{M}_{x x^{\prime} y y^{\prime}} = \sum_{ij}^{~} \left(U^{~}_{\rm X}\right)_{ix} M^{(n)}_{ijyy^{\prime}} \left(U^{~}_{\rm X}\right)_{jx^{\prime}} \, . 
\end{equation}
By contracting two tensors $\tilde{M}_{~}$ along the $x$ axis, we define 
\begin{equation}
N^{~}_{x x^{\prime} y y^{\prime}} = \sum_{s}^{~} \tilde{M}^{~}_{x s y^{~}_{1} y^{\prime}_{1}} \tilde{M}^{~}_{s x^{\prime} y_{2} y_{2}^{\prime}} \, ,
\end{equation}
where $y = y_1 \otimes y_2$ and $y^{\prime} = y^{\prime}_1 \otimes y^{\prime}_2$. 
We finally obtain the vertical projector $U^{~}_{\rm Y}$ when applying the SVD to the matrix unfolding $N^{\prime}_{y \, , \, \left(y^{\prime} x x^{\prime}\right)} = N^{~}_{x x^{\prime} y y^{\prime}}$ 
\begin{equation}
N^{\prime}_{~} = U^{~}_{\rm Y} \omega_{\rm Y} V^{\dag}_{\rm Y} \, . 
\end{equation}
Lastly, the second index dimension of $U^{~}_{\rm Y}$ is truncated down to the $D$.

\section{Numerical Results}
To study the critical behavior of the fractal Ising model in Fig.~\ref{fig:Fig_1}, we analyze the physical quantities of interest obtained locally by the technique of impurity tensor in Sec.~\ref{local_impurities_subsec} and globally by differentiating the free energy per site in Sec.~\ref{global_subsec}. 
The technique of impurity tensors is also used for obtaining the two-point correlation function in Sec.~\ref{corr_fun_subsec}.

The impurity tensor is a particular local tensor containing a local observable such as the Ising variable $\sigma_i$ at site $i$. 
Since the fractal lattice is a non-homogeneous system, one might expect a position dependence of the observation.
The site's location $i$ within the system is determined by the series of extensions of the impurity tensor. 
In our calculations, we aim to keep the observation site $i$ far from the ``outer'' boundary of the system.
More specifically, at each extension step, we insert a single impurity tensor alternatively to the upper left and the lower right site in the four sites in the center of the 12-cluster defined by Eq.~\eqref{T_ext}.

The complete information on the global behavior of the system is captured by the free energy
\begin{equation}
F_n = - k_{\rm B} T \ln{Z_n} \, .
\end{equation}
where $k_{\rm B}$ is the Boltzmann constant, which we set to one in the calculations $k_{\rm B} = 1$. 
Numerically, the convergence (\textit{i.e.} the thermodynamic limit) of the free energy per site 
\begin{equation}
f = \lim_{n \to \infty} \frac{F_n}{N_n} \, ,
\end{equation}
for the fractal lattice under study is already achieved for $n \sim 20$ as the number of sites grows exponentially as $N_n = 12^n$.

For convenience, let us list the most commonly used thermodynamic functions derived from $f$. 
The first derivative of the free energy with respect to temperature $T$ results in the {\em internal energy}
\begin{equation} \label{int_eng}
u = - T^2 \dfrac{\partial \left(f/T\right)}{\partial T}\, .
\end{equation}
The consequent temperature derivative of the internal energy yields the {\em specific heat}
\begin{equation} \label{spec_heat}
c = \dfrac{\partial u}{\partial T}=-T\dfrac{\partial^2 f}{\partial T^2}\, ,
\end{equation}
which has a non-analytic behavior at a phase transition.
Analogously, the first derivative of the free energy with respect to an external field $h$ results in the spontaneous magnetization
\begin{equation}
m = -\left. \dfrac{\partial f(h, T)}{\partial h} \right|_{h \to 0} \, , 
\end{equation}
and the second derivative of the free energy specifies the \textit{magnetic susceptibility}
\begin{equation}
\chi = \left. \dfrac{\partial m}{\partial h} \right|_{h \to 0} \, . 
\end{equation}
The application of automatic differentiation to a tensor network program for a fractal lattice is conceptually identical to the case of a square lattice as developed in Ref.~\cite{ad1}.
In both cases, the computation process, including the tensor network contractions, is represented as a directed acyclic graph, commonly referred to as a computation graph.
In their pioneering work, which is reported in Ref.~\cite{ad1}, the authors showcase the effectiveness of the differentiable programming tensor network approach through two applications: (1) computation of higher-order derivatives of the free energy and (2) gradient-based optimization of iPEPS.
The first application is relevant to our study as our focus is on obtaining derivatives of the partition function.
The primary technical distinction between our study and the techniques presented in Ref.\cite{ad1} lies in our usage of the HOTRG algorithm, specifically adapted for a fractal lattice, whereas Ref.\cite{ad1} employs TRG on a square lattice.
The use of HOTRG is deemed more suitable for the fractal lattice under investigation.
However, both HOTRG and TRG increase the lattice size exponentially, utilizing similar basic operations, including tensor index permutation, truncated singular value decomposition, and tensor contractions.
Like Ref.~\cite{ad1}, we adopt the reverse-mode automatic differentiation technique and incorporate the temperature $T$ as an input parameter. 
Furthermore, we include the external field $h$ as another input parameter.
To enhance numerical stability, we employ a custom linear algebra automatic differentiation library for truncated SVD, as described in Ref.~\cite{ad1}.

\subsection{Local impurities} \label{local_impurities_subsec}

%
\subsubsection{Exponent $\alpha$}
\begin{figure}[tb]
\includegraphics[width=0.48\textwidth,clip]{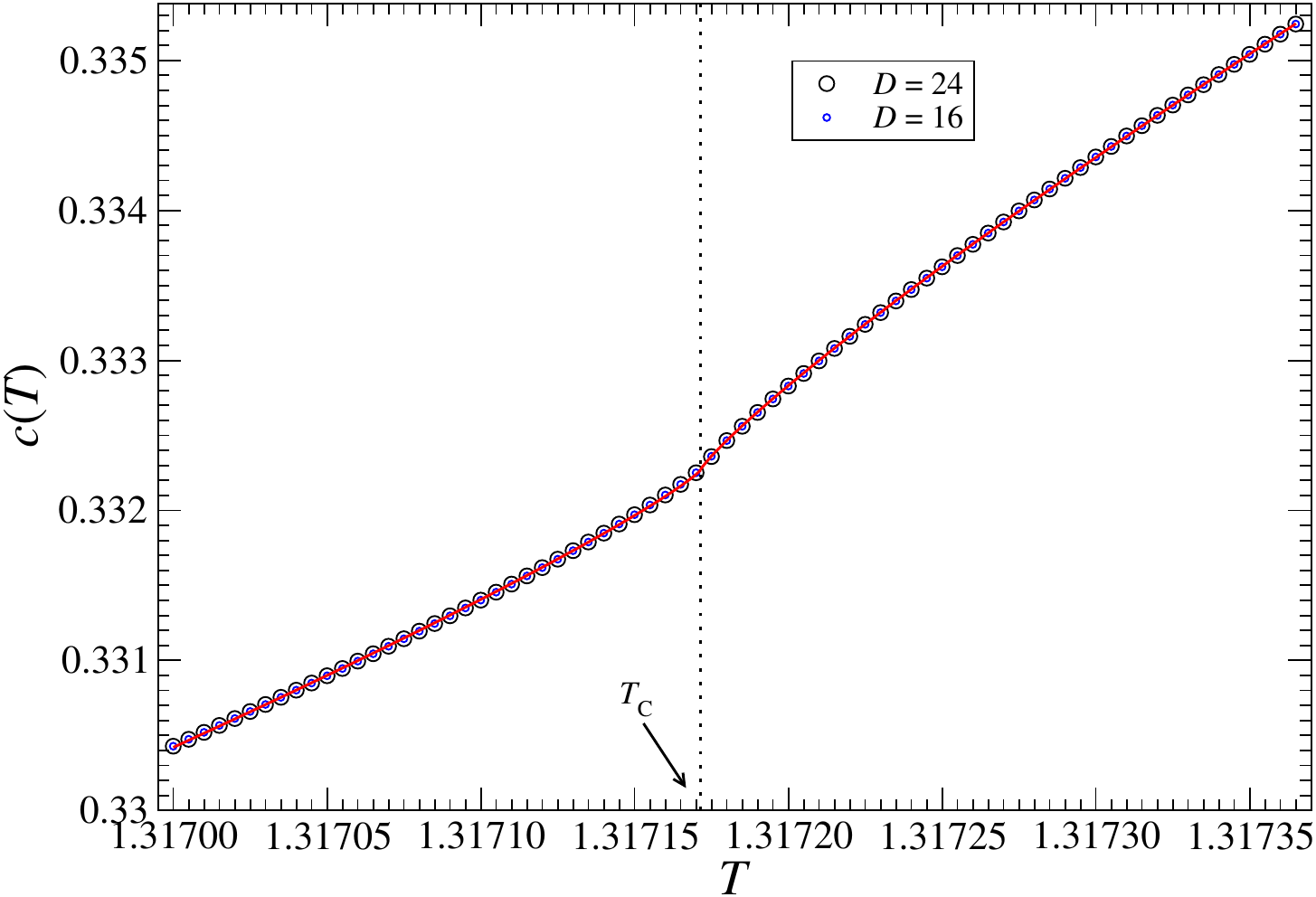}
\caption{
(Color online) 
Specific heat $c(T)$ as a function of temperature calculated from the (local) bond energy $u$ in the close vicinity of $T_{\rm c}^{~}$.
Specific heat $c$ was obtained by automatic differentiation as a first derivative of $u$ with respect to the temperature $T$. 
Data points with the bond dimension $D = 16$ and $D = 24$ are depicted as smaller blue circles and bigger black circles, respectively, whereas the fitting curve for $D = 24$ is depicted as a red line. 
Fitting yielded $\alpha \approx -0.87$ and $T_{\rm c}^{~} \approx 1.3171715$ (with $D = 24$).
A vertical dotted line indicates the critical temperature $T_{\rm c}^{~}$.
}
\label{fig:alpha_local1}
\end{figure}

\begin{figure}[tb]
\includegraphics[width=0.48\textwidth,clip]{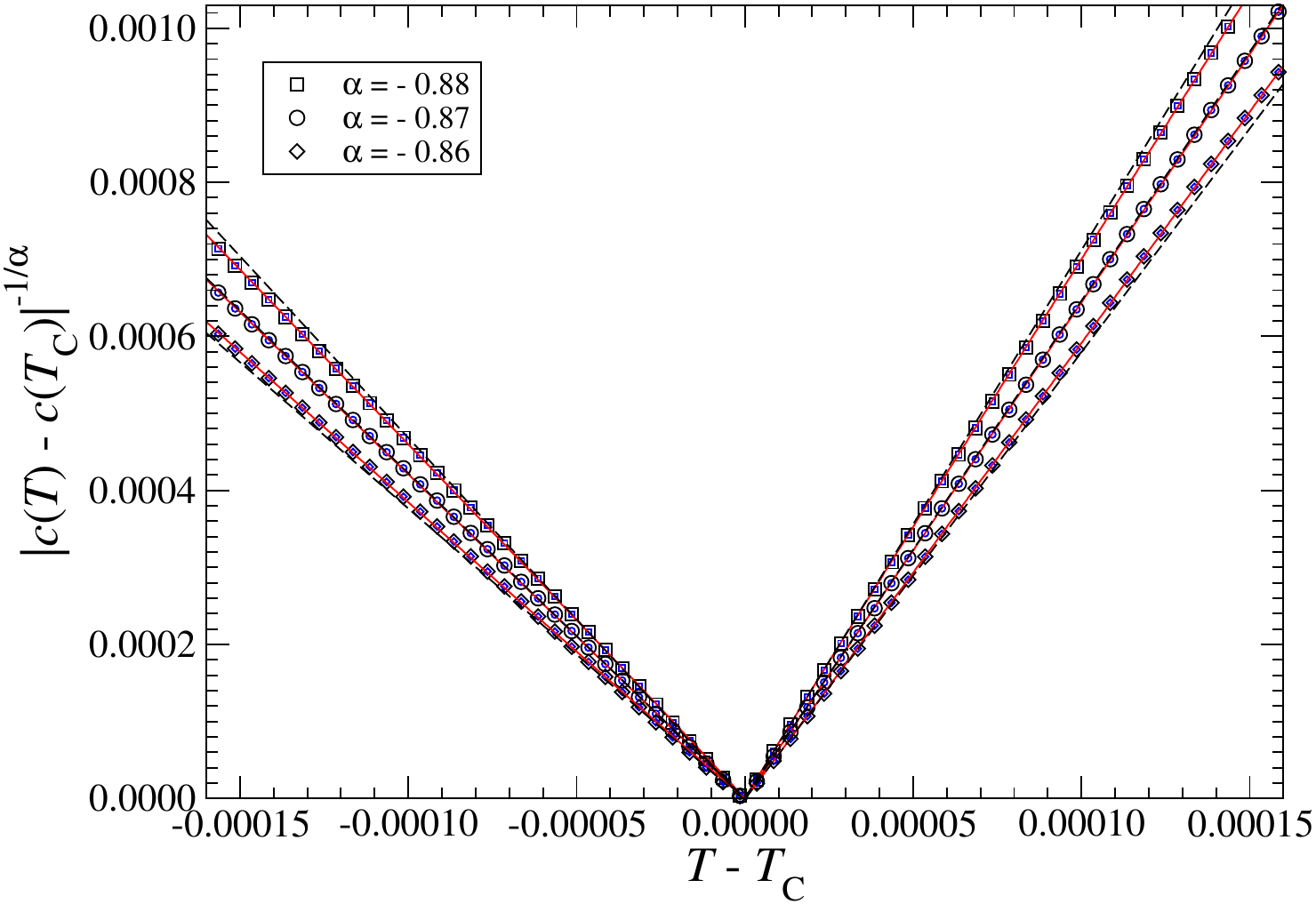}
\caption{ 
(Color online) 
The detail view of the linear dependence of $\left|c(T_{~}^{~}) - c(T_{\rm c}^{~})\right|^{-1/\alpha}$ with respect to the temperature near $T_{\rm c}^{~}$, where $c$ is the local specific heat. 
Three values of $\alpha$ are presented: (1) $\alpha = -0.88$ (smaller than our estimate of $\alpha$, shown as squares), (2) $\alpha = -0.87$ (corresponding to our estimate, show as circles), and (3) $\alpha = -0.86$ (larger than our estimate of $\alpha$, show as diamonds). 
Data points with the bond dimension $D = 16$ and $D = 24$ are depicted as smaller blue and bigger black shapes, respectively, whereas the fitting curve for $D = 24$ is depicted as a red line. 
}
\label{fig:alpha_local2}
\end{figure}

The specific heat $c(T)$  does not diverge at the critical temperature $T_{\rm c}^{~}$ as shown in Refs.~\cite{2dising, j1j2}.
Instead, $c(T)$ exhibits a weak non-analytic behavior at $T \approx T_{\rm c}^{~}$ with its broadened maximum shifted to $T > T_{\rm c}^{~}$.
However, a numerical derivative of $c(T)$ with respect to temperature has a sharp peak at $T_{\rm c}^{~}$, \textit{i.e.}, $T_{\rm c}^{~} = \max_{T} \left\{d c(T) / dT \right\}$.
Here, the specific heat $c$ was obtained by automatic differentiation from the (local) bond energy $c = \partial u / \partial T$, where $u$ is calculated using impurity tensor which contains a single bond spin-spin correlation term $\sigma_i \sigma_j$ of two neighboring sites $i$ and $j$ located somewhere far from the external surface of the fractal lattice. 
Fitting the data points in the vicinity of $T_{\rm c}^{~}$ (as estimated from the spontaneous magnetization) to the form
\begin{equation} \label{beta_exp_def}
c\left(T\right) \propto |T - T_{\rm c}^{~}|^{-\alpha} \, ,
\end{equation}
we obtained the value of the critical exponent $\alpha \approx -0.87$ and $T_{\rm c}^{~} \approx 1.3171715$ using the bond dimension $D = 24$, see Fig.~\ref{fig:alpha_local1}. 
The relative difference in the estimate of $\alpha$ between the numerical result with $D = 16$ and $D = 24$ is less than $0.1\%$. 
The precision of the estimate of the exponent $\alpha$ can be also judged from a tiny deviation from the linear dependence (the dashed lines) in $\left|c(T_{~}^{~}) - c(T_{\rm c}^{~})\right|^{-1/\alpha}$ near $T_{\rm c}^{~}$, see Fig.~\ref{fig:alpha_local2}; setting the value of the exponent $\alpha$ slightly below or slightly above the value $\alpha = -0.87$ results in a visibly non-linear behavior of $\left|c(T_{~}^{~}) - c(T_{\rm c}^{~})\right|^{-1/\alpha}$ near $T_{\rm c}^{~}$.
%

%
\subsubsection{Exponent $\beta$}
Local magnetization $m$ was already calculated using the technique of impurity tensor inserted somewhere far from the external surface on the fractal lattice~\cite{2dising, APS}.
Fitting the data points at $T \lesssim T_{\rm c}^{~}$ according to
\begin{equation}
m(T) \propto \left(T_{\rm c}^{~} - T\right)_{~}^{\beta} \, , \\
\end{equation}
$\beta$ was found to be $\beta \approx 0.01388$ and $T_{\rm c}^{~} \approx 1.31717$ with $D = 32$ in Ref.~\cite{APS}.
Our new estimate, obtained by fitting the magnetization calculated very close to $T_{\rm c}^{~}$ with a very fine step ($\Delta T = 5 \times 10^{-6}$), is $\beta \approx 0.01383$ and $T_{\rm c}^{~} = 1.3171724$ with $D = 24$.
The numerical results do not change much when increasing the bond dimension above $D=16$; we checked the relative difference in the estimate of $\beta$ between $D = 16$ and $D = 24$ is less than $0.1\%$. 

%
\subsubsection{Exponent $\gamma$}
\begin{figure}[tb]
\includegraphics[width=0.48\textwidth,clip]{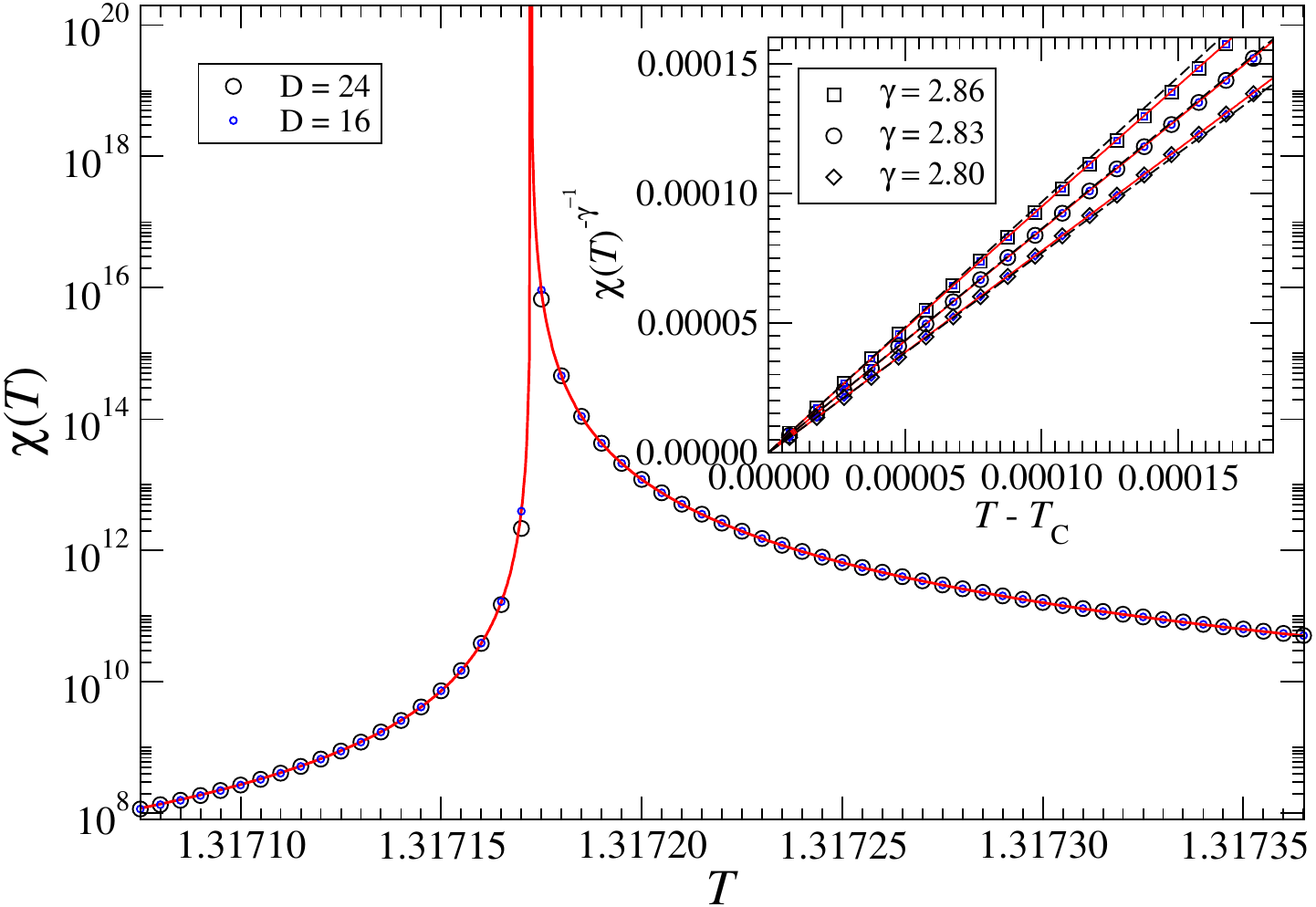}
\caption{ 
(Color online) 
Magnetic susceptibility $\chi(T)$ as a function of temperature $T$ in the vicinity of the critical temperature $T_{\rm c}^{~}$. 
Magnetic susceptibility $\chi$ was obtained by automatic differentiation as the first derivative of (local) magnetization $m$ with respect to the global field $h$. 
Data points with the bond dimension $D = 16$ and $D = 24$ are depicted as smaller blue circles and bigger black circles, respectively, whereas the fitting curve for $D = 24$ is depicted as a red line.
Inset: The linear dependence of $\chi(T)^{-1/\gamma}$ above $T_{\rm c}^{~}$. 
}
\label{fig:gamma_local}
\end{figure}

Magnetic susceptibility $\chi$ was obtained by automatic differentiation as a derivative of local magnetization $m$ with respect to the global field $h$, i.e. $\chi = \partial m / \partial h$. 
Fitting the data points (with $D = 24$) around $T_{\rm c}^{~}$ according to 
\begin{equation}
\chi(T) \propto |T - T_{\rm c}^{~}|^{-\gamma} \, , 
\end{equation}
we obtained $T_{\rm c}^{~} \approx 1.3171723$ and slightly different values of $\gamma$ below and above $T_{\rm c}^{~}$, namely $\gamma^{-} \approx 2.797$ and $\gamma^{+} \approx 2.826$, respectively. 
We consider $\gamma^{+}$ to be more accurate than $\gamma^{-}$ since the power-law behavior appears to be more reliable above $T_{\rm c}^{~}$. 
Thus, we assume $\gamma = \gamma^{+}$.
The relative difference in $\gamma^{+}$ between $D = 16$ and $D = 24$ is less than $0.1\%$, whereas the relative difference in $\gamma^{-}$ between $D=16$ and $D=24$ is around $0.5\%$. 
Magnetic susceptibility $\chi$ with $D = 16$ and $D = 24$ is presented in Fig.~\ref{fig:gamma_local}, where the inset clearly shows the linear dependence of $\chi(T)^{-1/\gamma}$ when $T \gtrsim T_{\rm c}^{~}$. 
Four data points very close to $T_{\rm c}^{~}$ were excluded from the numerical analysis since there is a visible difference in the magnetic susceptibility between $D=16$ and $D=24$, see Fig.~\ref{fig:gamma_local} (namely we excluded $T = \left\{1.317165, 1.317170, 1.317175, 1.317180\right\}$). 
%

%
\subsubsection{Exponent $\delta$}
The magnetic field response at the critical temperature $T=T_{\rm c}^{~}$
\begin{equation}
m(h, T=T_{\rm c}^{~}) \propto h^{1/\delta} \quad \text{as} \quad  h \to 0,
\end{equation}
was analyzed in Ref.~\cite{APS}, where the associated critical exponent $\delta$ was found to be $\delta \approx 206$ with a relatively low value of the bond dimension $D=12$. 
To determine the critical exponent $\delta$ more accurately, we reproduced the previous calculations of the local magnetization as a function of (small) global magnetic field $h$ ($2 \times 10^{-8} \leq h \leq 5 \times 10^{-7}$) at the critical point $T = T_{\rm c}^{~} = 1.3171724$ using larger bond dimension. 
Our new and much more reliable estimate is $\delta \approx 204.93$ with $D=24$; the difference in $\delta$ between $D=16$ and $D=24$ was negligible.

\subsection{Two-point correlation function and critical exponents $\nu$ and $\eta$} \label{corr_fun_subsec}

Here we calculate the correlation between two (equivalent) spin variables $s_0^{~}$ and $s_r^{~}$ as a function of the distance $r$ 
\begin{equation} \label{corr}
G(r) = \left\langle s_0^{~} s_r^{~} \right\rangle,
\end{equation}
where the distance $r$ grows exponentially as $r=\left\{ 4^0, 4^1, 4^2, 4^3, \cdots, 4^{n} \right\}$ with the number of the iteration steps $n$ due to the fixed growth process of the fractal lattice.
We calculate the correlation $G(r)$ using the technique of impurity tensor in three steps.
(1) We start with the single-point impurity (the same as when calculating local magnetization), which is extended up to the required size. 
(2) Next, we merge two single-point impurities into one two-point impurity. 
The distance $r$ is given by the location of the spin variables $s_0^{~}$ and $s_r^{~}$ at the merge step. 
(3) Finally, we simply extend the two-point impurity tensor until the correlation $G(r)$ converges numerically. 
Our technique is an adaptation of the technique based on TRG proposed in Ref.~\cite{TRG} for translationally invariant systems.
The numerical calculations of the correlation function for the two-dimensional classical Ising model using TRG were performed in Ref.~\cite{Correlations}.

\begin{figure}[tb]
\includegraphics[width=0.48\textwidth,clip]{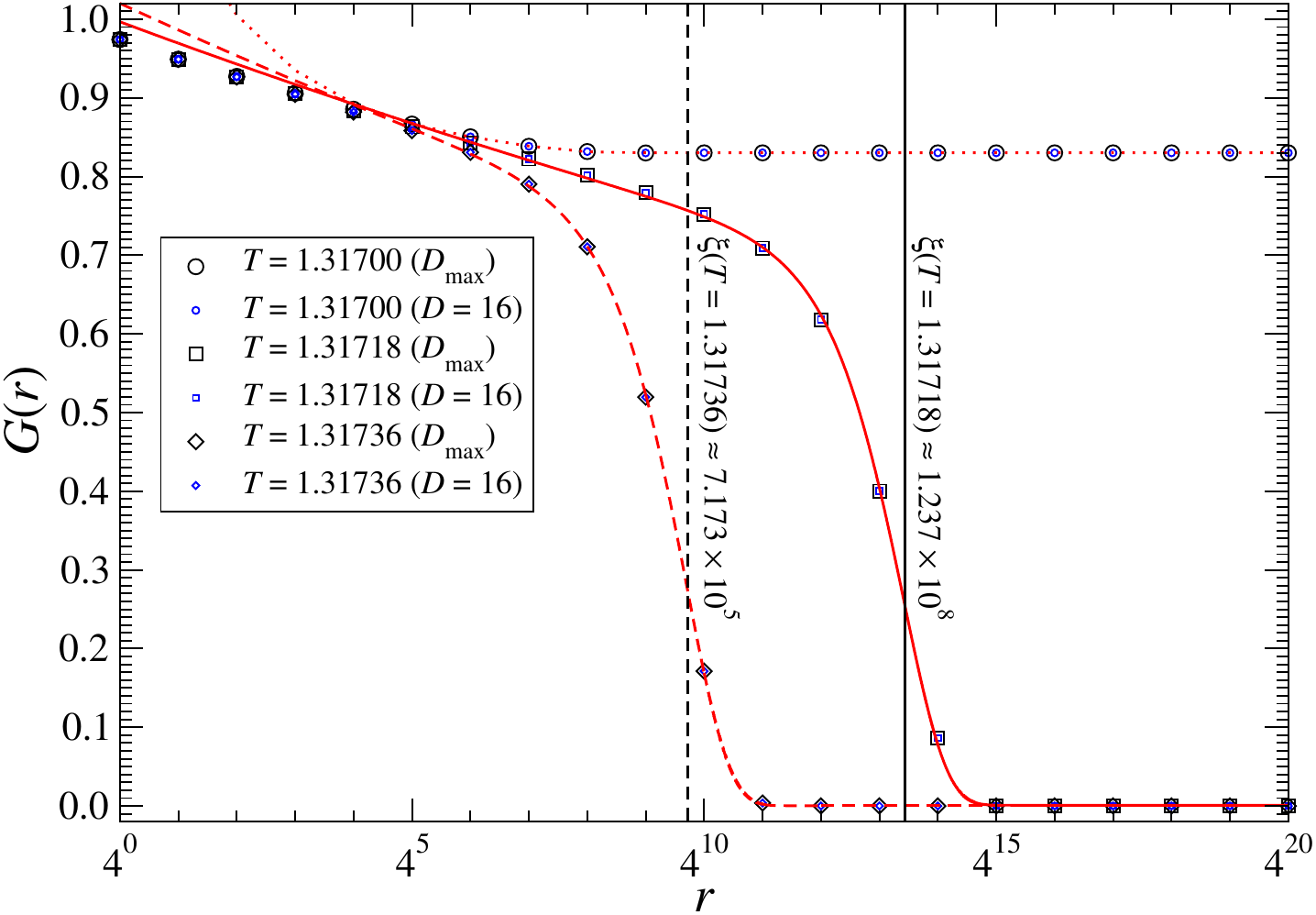}
\caption{
(Color online) 
The correlation $G(r)$ as a function of distance $r$. 
The correlation $G(r)$ for three different temperatures is shown: 
(1) $T = 1.317$ (below $T_{\rm c}^{~}$), 
(2) $T = 1.31718$ (very close to $T_{\rm c}^{~}$), 
and (3) $T = 1.31736$ (above $T_{\rm c}^{~}$).
Here, the bond dimension was implemented adaptively where the normalized singular values smaller than $10^{-14}$ were discarded.
$D_{\rm max}^{~}$ denotes the maximal (unbounded) adaptive dimension (in practice $D_{\rm max}^{~} \leq 23$).
We present data points calculated with $D_{\rm max}^{~}$ and $D=16$, which we depict as black and blue symbols, respectively. 
The fitting curves Eq.~\eqref{gr_acexi} were obtained for $D_{\rm max}^{~}$ and are depicted by red lines.
The values of the correlation lengths $\xi$ for $T = 1.31718$ and $T = 1.31736$ are indicated by full and dashed line, respectively. 
The numerical analysis did not include data points for $r < 1024$. 
}
\label{fig:corr1}
\end{figure}

It is expected that near the critical temperature $T \approx T_{\rm c}^{~}$, the correlation function
\begin{equation} \label{corr_func}
\Gamma (r) = \left\langle s_0^{~} s_r^{~} \right\rangle - \left\langle s_0^{~} \right\rangle \left\langle s_r^{~} \right\rangle
\end{equation}
decays exponentially with the distance $r$ (for $r \rightarrow \infty$)
\begin{equation} \label{gr_decay}
\Gamma (r) \propto r^{-E} e^{-r / \xi} \, ,
\end{equation}
where $\xi$ is the correlation length and $E$ is some number which is equal to $\left(d - 2 + \eta\right)$ at the critical temperature $T = T_{\rm c}^{~}$ ($d$ being the spatial dimension and $\eta$ is the critical exponent associated with the correlation function).
Therefore, we assume the fitting function to be of the form
\begin{equation} \label{gr_acexi}
G(r) = C^2_{~} + \dfrac{A^{~}_{~}}{r^{E}_{~}} \exp(-r/\xi) \, , 
\end{equation}
where $A$, $C$, $E$, and $\xi$ are the fitting parameters we want to estimate at each temperature independently.
Comparing the last four equations, one can see that $C$ is the magnetization, and this observation is indeed consistent with our numerical results. 
The correlation $G(r)$ as a function of the distance $r$ is presented in Fig.~\ref{fig:corr1} in the case of three temperature regimes: 
(1) below $T_{\rm c}^{~}$ at $T = 1.317$, 
(2) very close to the critical temperature $T_{\rm c}^{~}$ at $T = 1.31718$, 
and (3) above $T_{\rm c}^{~}$ at $T = 1.31736$.
In our numerical calculations of $G(r)$, the bond dimension $D$ was implemented adaptively where the normalized singular values smaller than a certain threshold were discarded (we used as the threshold $\varepsilon = 10^{-14}$), which significantly reduced the computational time; however, the results are indistinguishable from faithful fixed-bond dimension calculations.
The maximal (unbounded) adaptive dimension achieved was $D=23$, which we denote as $D_{\rm max}^{~}$.
To verify the convergence with respect to the bond dimension cut, we compare the data points obtained for $D_{\rm max}^{~}$ and $D=16$ in Fig.~\ref{fig:corr1}, which we depict as black and blue symbols, respectively. 
The fitting curves according to Eq.~\eqref{gr_acexi} were obtained with $D_{\rm max}^{~}$ and are depicted by red lines.
The values of the correlation lengths $\xi$ for $T = 1.31718$ and $T = 1.31736$ are indicated by full and dashed line, respectively. 
Data points for small values of $r$ (\textit{i.e.}, $r<1024$) were not included in the numerical analysis as we care for large values of $r$ only ($r \rightarrow \infty$). 
For concreteness, as for this particular example, we obtained the following values with the bond dimension $D_{\rm max}^{~}$: 
(1) $E \approx 0.369$, $\xi \approx 3.61 \times 10^{4}$, $C \approx 0.91$ for $T = 1.31700$,
(2) $E \approx 0.02$, $\xi \approx 1.237 \times 10^{8}$, $C \approx 0.03$ for $T = 1.31718$, and
(3) $E \approx 0.024$, $\xi \approx 7.173 \times 10^{5}$, $C \approx 0.016$ for $T = 1.31736$.

\begin{figure}[tb]
\includegraphics[width=0.48\textwidth,clip]{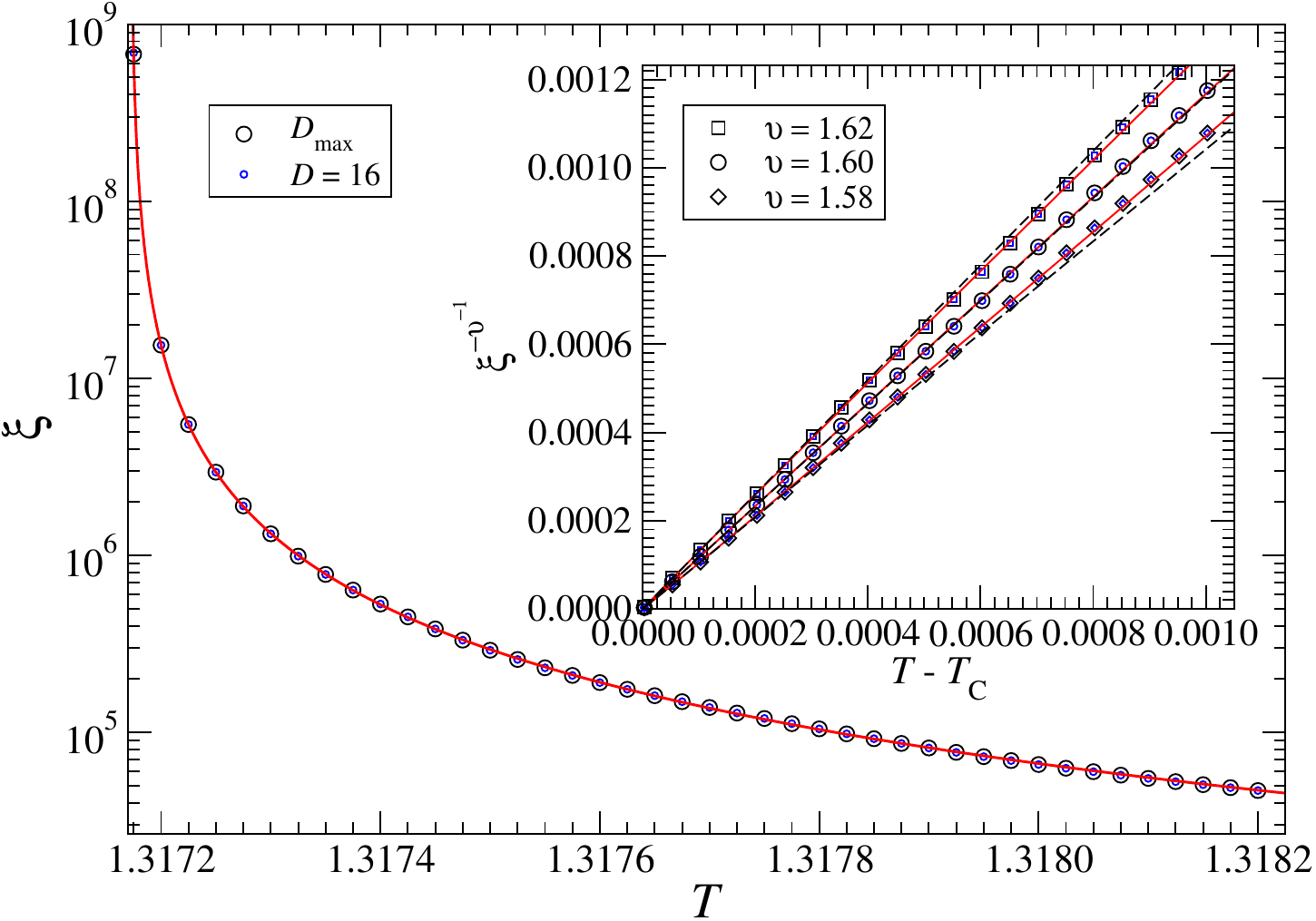}
\caption{
(Color online) 
The correlation length $\xi$ as a function of temperature $T$ above $T_{\rm c}^{~}$. 
The data points were obtained by fitting the correlation function $G(r)$ for $r\geq 1024$ at each temperature $T$ separately. 
The fitting curve was obtained for $D_{\rm max}^{~}$ and is depicted by a red line. 
The resulting value of exponent $\nu$ was found to be $\nu \approx 1.60$ and $T_{\rm c}^{~} \approx 1.3171723$. 
Inset: Critical scaling of $\xi(T)^{\nu^{-1}}$ above the critical temperature $T_{\rm c}^{~}$.
}
\label{fig:nu}
\end{figure}
%

%
\subsubsection{Exponent $\nu$}
The critical exponent $\nu$ and the critical temperature $T_{\rm c}^{~}$ can be obtained by fitting the correlation length $\xi(T)$ as a function of temperature in the vicinity of $T_{\rm c}^{~}$ according to (see Fig.~\ref{fig:nu})
\begin{equation} \label{nu_exp_def}
\xi(T) \propto \left|T- T_{\rm c}^{~}\right|^{-\nu} \, .
\end{equation}
The data points $\xi(T)$ were obtained by fitting the correlation $G(r)$ for $r\geq 1024$ according to the Eq.~\eqref{gr_acexi} for each temperature $T$ separately. 
The correlation length $\xi(T)$ below $T_{\rm c}^{~}$ exhibits a slightly unstable scaling behavior; therefore, we will omit the numerical analysis below $T_{\rm c}^{~}$ here. 
The fitting curve obtained for $D_{\rm max}^{~}$ is depicted by a red line in Fig.~\ref{fig:nu}. 
The resulting value of $\nu$ was found to be approximately $\nu \approx 1.60$ and $T_{\rm c}^{~} \approx 1.3171723$ (both values being independent of $D$). 
The precision of the estimate of $\nu$ can be assessed from a tiny deviation from the linear dependence (the dashed lines) in $\xi(T)^{-1/\nu}$ near $T_{\rm c}^{~}$, see inset of Fig.~\ref{fig:nu}; setting the value of the exponent $\nu$ slightly below or slightly above the value $\nu = 1.60$ results in a visibly non-linear behavior.
\begin{figure}[tb]
\includegraphics[width=0.48\textwidth,clip]{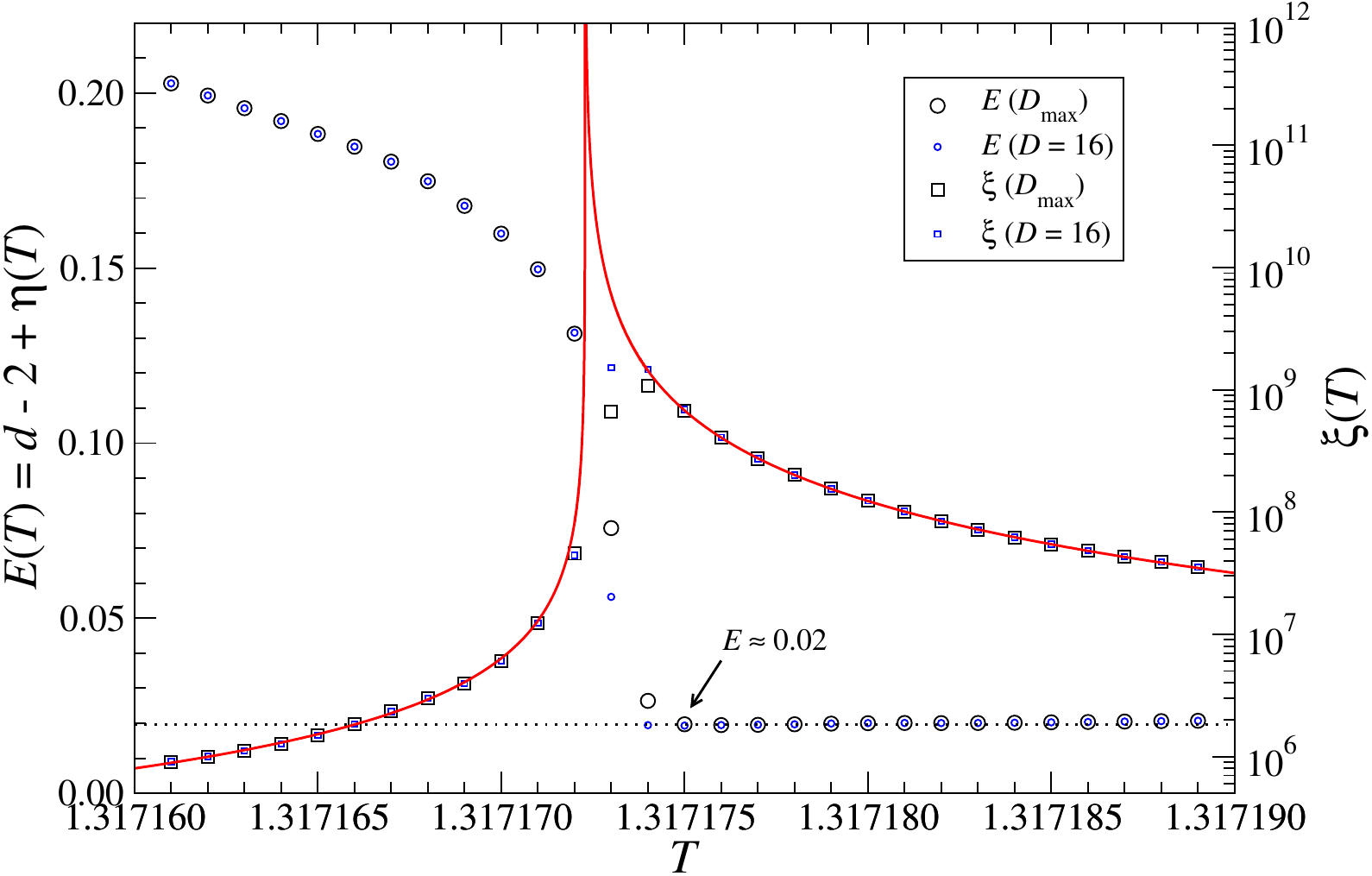}
\caption{
(Color online) 
A detailed view of the exponent $E(T)$ and the correlation length $\xi(T)$ as functions of temperature $T$ in the vicinity of $T_{\rm c}^{~}$. 
The correlation length $\xi(T)$ is presented here to indicate the location of the critical temperature $T_{\rm c}^{~}$. 
The value of the exponent $E(T)$ seems to be nearly constant (indicated by a dotted horizontal line) in the right vicinity of $T_{\rm c}^{~}$, where we observed $E(T) \approx 0.02$. 
}
\label{fig:eta}
\end{figure}
%

%
\subsubsection{Exponent $\eta$}
Let us now turn our attention to the exponent $E$, which appears in the Eqns.~(\ref{gr_decay},~\ref{gr_acexi}). 
At the critical temperature $T_{\rm c}^{~}$, we know that $E$ can be expressed as a sum of the spatial dimension $d$ and the critical exponent $\eta$, \textit{i.e.}, $E = d - 2 + \eta$. 
Of course, calculating $G(r)$ directly at $T_{\rm c}^{~}$ and then estimating the value of $E(T_{\rm c}^{~})$ would be heavily affected by the truncation error.
Luckily, it seems that $E(T)$ is nearly stable for $T \gtrsim T_{\rm c}^{~}$ taking the value $E \approx 0.02$, see Fig.~\ref{fig:eta}. 
We consider as our effective estimate of $E(T_{\rm c})$ to be the first point above true $T_{\rm c}^{~}$ where $E^{\left[D=16\right]}(T) \approx E(T)^{\left[D=D_{\rm max}^{~}\right]}$ (which is found to be around $T = 1.317175$).
Below $T_{\rm c}$, the value of $E(T)$ changes more rapidly; however it seems to decay to the value $E \approx 0.02$ observed above $T_{\rm c}$. 
As one can expect, there are significant differences in the estimates of $\xi$ and $E$ between $D = 16$ and $D_{\rm max}^{~}$ at $T \approx T_{\rm c}$ due to the finite bond dimension cut, see Fig.~\ref{fig:eta}; however, these differences become negligible when moving away from $T_{\rm c}$ by a tiny step.

\subsection{Global behavior} \label{global_subsec}

%
\subsubsection{Exponent $\alpha$}
\begin{figure}[tb]
\includegraphics[width=0.48\textwidth,clip]{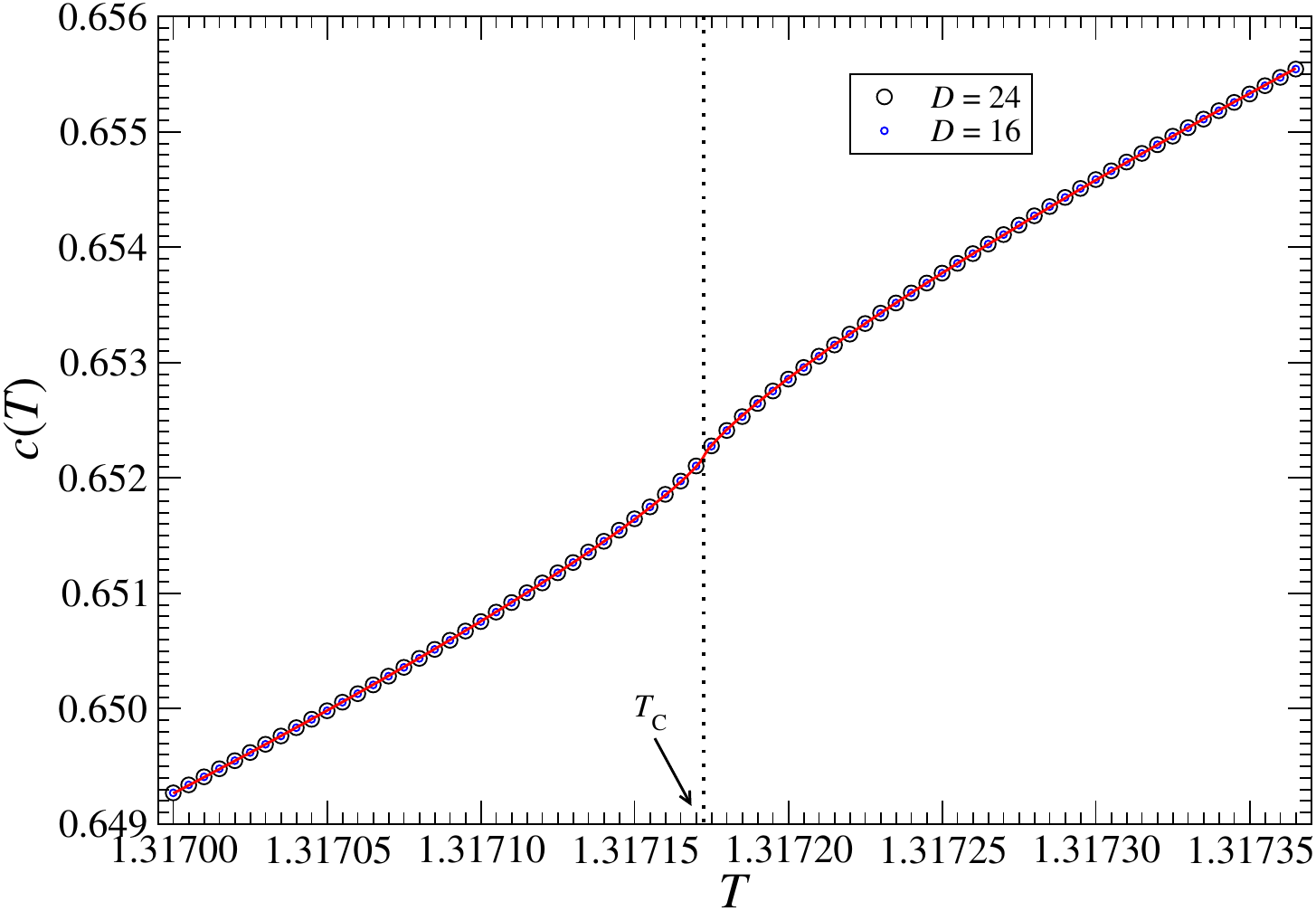}
\caption{ 
(Color online) 
A detailed view of the global specific heat $c(T)$ in the vicinity of the critical temperature $T_{\rm c}^{~}$. 
Global specific heat $c(T)$ is calculated as a second derivative of the free energy per site $f(T)$ with respect to the temperature by the automatic differentiation.
Data points for the bond dimension $D=24$ are depicted as big black circles, whereas for $D=16$ as smaller blue circles. 
The fitting curve is obtained for $D=24$ and is shown as a thick red line. 
For $D=24$, the fitting yielded $\alpha = - 0.824$ and $T_{\rm c}^{~} = 1.3171725$.
A vertical dotted line indicates the location of the critical temperature $T_{\rm c}^{~}$.
}
\label{fig:alpha_global1}
\end{figure}

\begin{figure}[tb]
\includegraphics[width=0.48\textwidth,clip]{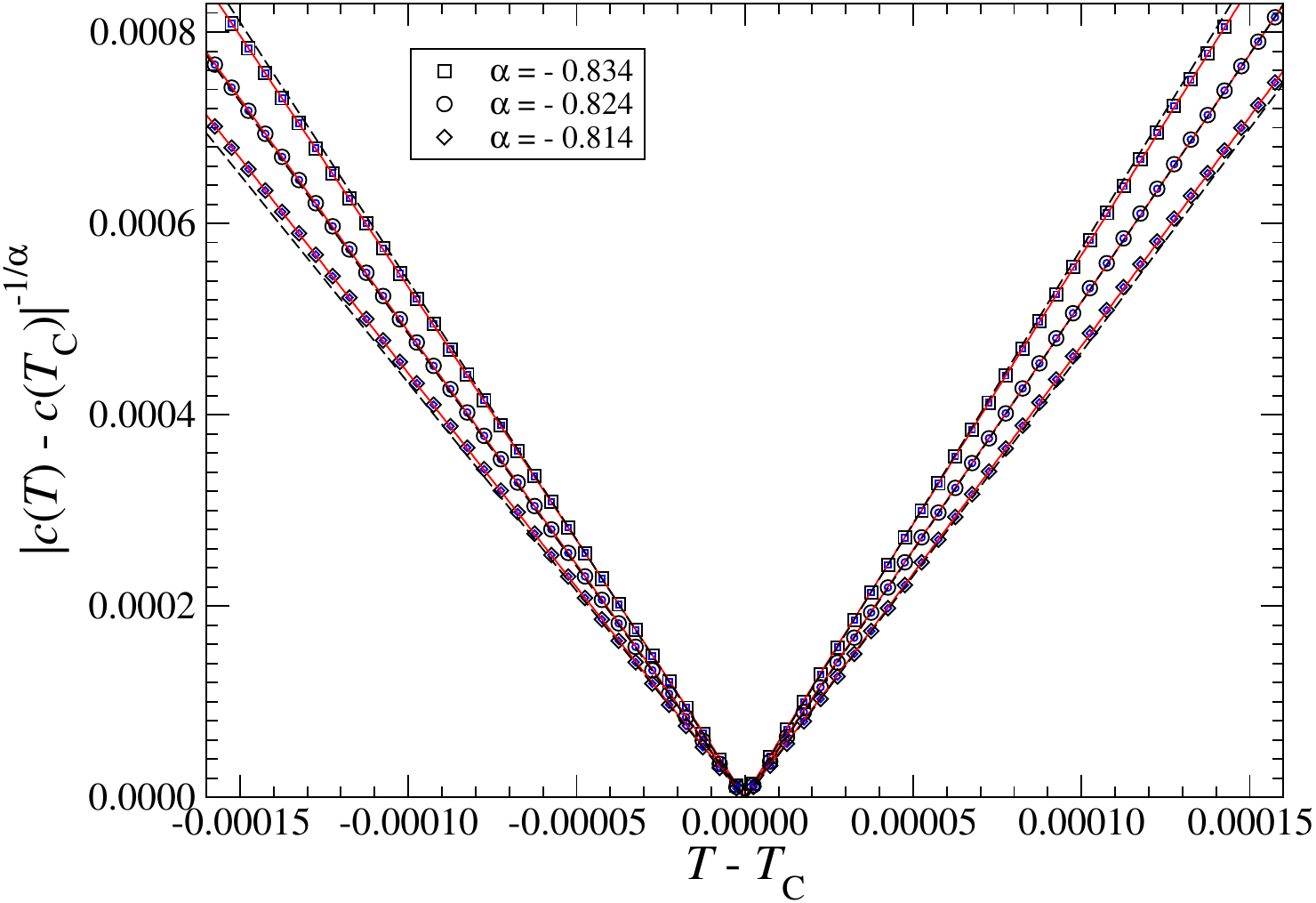}
\caption{ 
(Color online) 
The detail view of the linear dependence of $\left|c(T_{~}^{~}) - c(T_{\rm c}^{~})\right|^{-1/\alpha}$ with respect to the temperature near $T_{\rm c}^{~}$, where $c$ is the global specific heat. 
Three values of $\alpha$ are compared: (1) $\alpha = -0.834$ (smaller than our estimate of $\alpha$, shown as squares), (2) $\alpha = -0.824$ (corresponding to our estimate, shown as circles), and (3) $\alpha = -0.814$ (larger than our estimate of $\alpha$, shown as diamonds). 
Data points with the bond dimension $D = 16$ and $D = 24$ are depicted as smaller blue and bigger black shapes, respectively, whereas the fitting curve for $D = 24$ is depicted as a red line. 
}
\label{fig:alpha_global2}
\end{figure}

The global specific heat $c$ was obtained from the free energy per site $f$ by automatic differentiation as a second derivative with respect to the temperature, \textit{i.e.}, $c = - T \partial^2 f / \partial T^2$ (while keeping the magnetic field constant $h = 0$). 
Fitting the data points in the vicinity of the critical temperature $T_{\rm c}^{~}$, we obtained the value of the global critical exponent $\alpha \approx -0.824$ and $T_{\rm c}^{~} \approx 1.3171725$ with $D=24$, see Fig.~\ref{fig:alpha_global1}. 
As there is very good agreement between $D=16$ and $D=24$, no data points were excluded from the numerical analysis. 
The relative difference in the estimate of $\alpha$ between $D=16$ and $D=24$ is even smaller than in the case of the local exponent $\alpha$ extracted from the local impurities.
The precision of the estimate of the exponent $\alpha$ can be also assessed from a tiny deviation from the linear dependence (the dashed lines) in $\left|c(T_{~}^{~}) - c(T_{\rm c}^{~})\right|^{-1/\alpha}$ near $T_{\rm c}^{~}$, see Fig.~\ref{fig:alpha_global2}. 
%

%
\subsubsection{Exponent $\beta$}
\begin{figure}[tb]
\includegraphics[width=0.48\textwidth,clip]{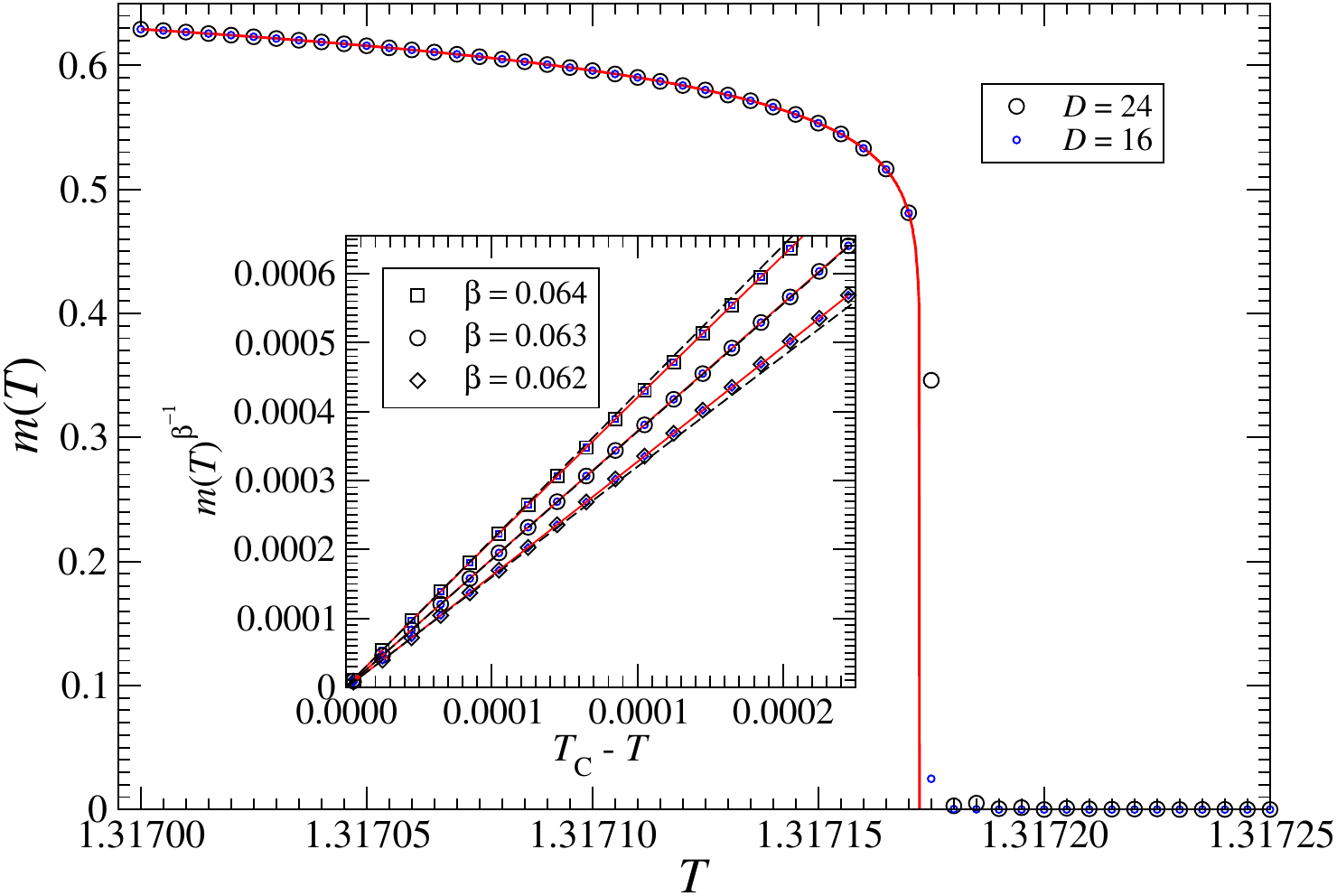}
\caption{
(Color online) 
A detailed view of the global spontaneous magnetization per site $m(T)$ in the vicinity of the critical temperature $T_{\rm c}^{~}$.
The global spontaneous magnetization per site $m(T)$ is calculated as a first derivative of the free energy $f(h, T)$ per site with respect to the external field $h$ using the automatic differentiation. 
The data points for the bond dimension $D=24$ and $D=16$ are depicted as big black circles and smaller blue circles, respectively. 
The fitting curve is obtained for $D=24$ and is shown as a thick red line here.
For $D=24$, the fitting yields $\beta = 0.0629$ and $T_{\rm c}^{~} = 1.3171724$. 
Inset: Critical scaling of $m(T)^{\beta^{-1}}$ below the critical temperature $T_{\rm c}^{~}$.
}
\label{fig:beta_global}
\end{figure}

The global magnetization $m$ was calculated by the automatic differentiation as a first derivative of the free energy per site $f(h, T)$ with respect to the external field $h$ (at exactly $h=0$), \textit{i.e.} $m = - \partial f(h, T) / \partial h$.
The numerical analysis was performed up to the temperature $T \leq 1.31717$ (including) since the magnetization for $D=16$ and $D=24$ is nearly indistinguishable up to that point.
For $D=24$, the fitting yields $\beta = 0.0629$ and $T_{\rm c}^{~} = 1.3171724$, see Fig.~\ref{fig:beta_global}. 
The numerical results do not change much when increasing the bond dimension above $D=16$; the relative difference in the estimate of $\beta$ between $D = 16$ and $D = 24$ is negligible. 
The linear dependence of $m(T)^{\beta^{-1}}$ below the critical temperature $T_{\rm c}^{~}$ is shown in the inset of Fig.~\ref{fig:beta_global}.
%

%
\subsubsection{Exponent $\gamma$}
\begin{figure}[tb]
\includegraphics[width=0.48\textwidth,clip]{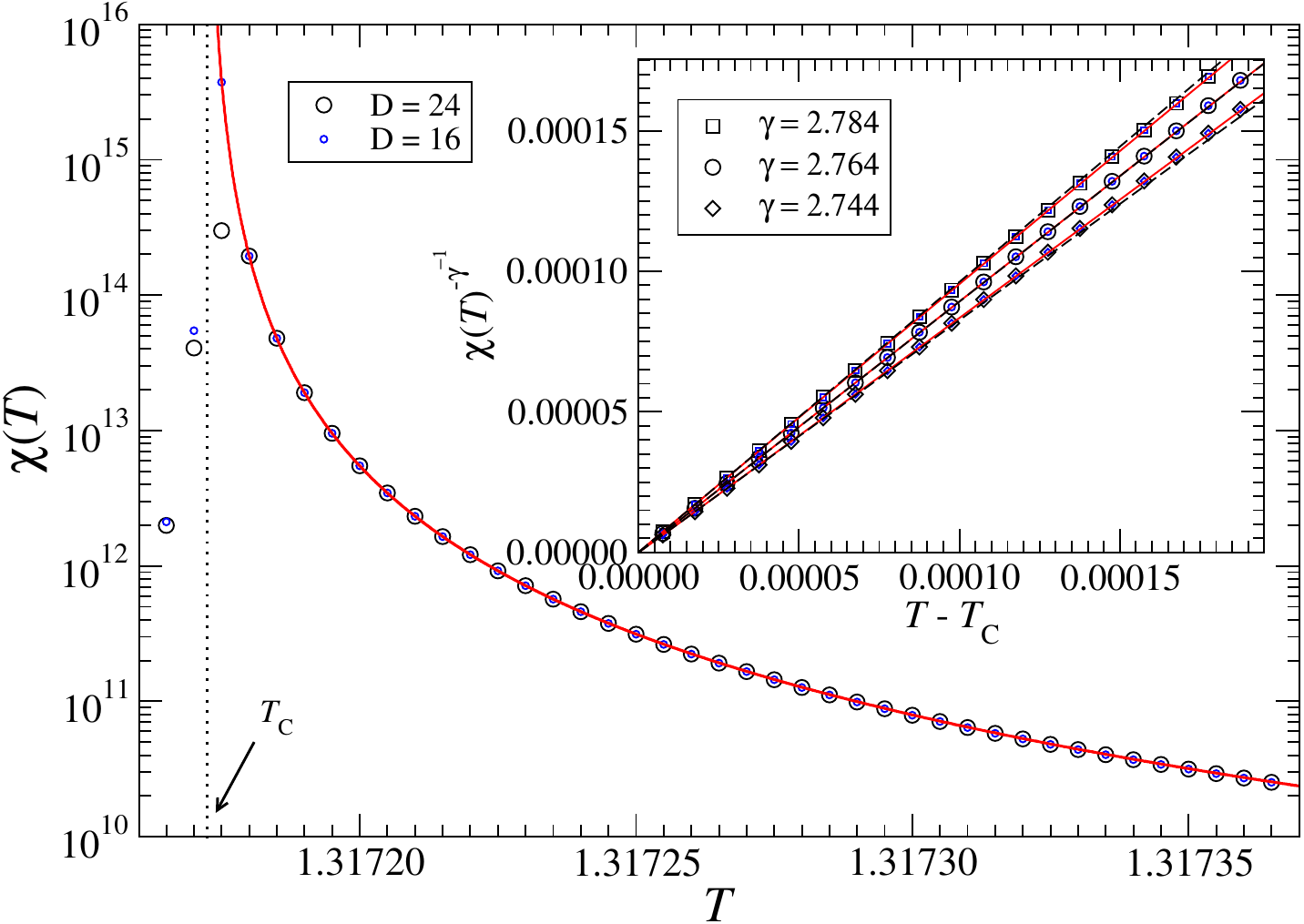}
\caption{
(Color online) 
A detailed view of the global magnetic susceptibility per site $\chi(T)$.
The global magnetic susceptibility per site $\chi(T)$ is calculated as a second derivative of the free energy per site $f(h, T)$ with respect to the external field $h$ using automatic differentiation.
The data points for the bond dimension $D=24$ and $D=16$ are depicted as big black circles and smaller blue circles, respectively. 
The fitting curve is obtained for $D=24$ and is shown as a thick red line here. 
For $D=24$, the fitting yields $T_{\rm c}^{~} = 1.3171724$ and $\gamma = 2.764$. 
A vertical dotted line indicates the location of the critical temperature $T_{\rm c}^{~}$.
Inset: Critical scaling of $\chi(T)^{-\gamma^{-1}}$ above the critical temperature $T_{\rm c}^{~}$.
}
\label{fig:gamma_global}
\end{figure}

The global magnetic susceptibility per site $\chi(T)$ is calculated as a second derivative of the free energy per site $f(h, T)$ with respect to the external field $h$ using the automatic differentiation (at exactly $h=0$), \textit{i.e.}, $\chi(T) = - \partial^2 f(h, T) / \partial h^2$.
The numerical analysis was performed only above $T_{\rm c}^{~}$, because the second derivative below $T_{\rm c}^{~}$ was unstable, and we also excluded the data points for $T \leq 1.31718$ to avoid fitting too close to $T_{\rm c}^{~}$.
For $D=24$, the fitting yields $\gamma = 2.764$ and $T_{\rm c}^{~} = 1.3171724$, see Fig.~\ref{fig:gamma_global}. 
The relative difference in $\gamma$ between $D=16$ and $D=24$ is less than $0.1\%$. 
The linear dependence of $\chi(T)^{-\gamma^{-1}}$ above the critical temperature $T_{\rm c}^{~}$ is shown in the inset of Fig.~\ref{fig:gamma_global}.
%

%
\subsubsection{Exponent $\delta$}
\begin{figure}[tb]
\includegraphics[width=0.48\textwidth,clip]{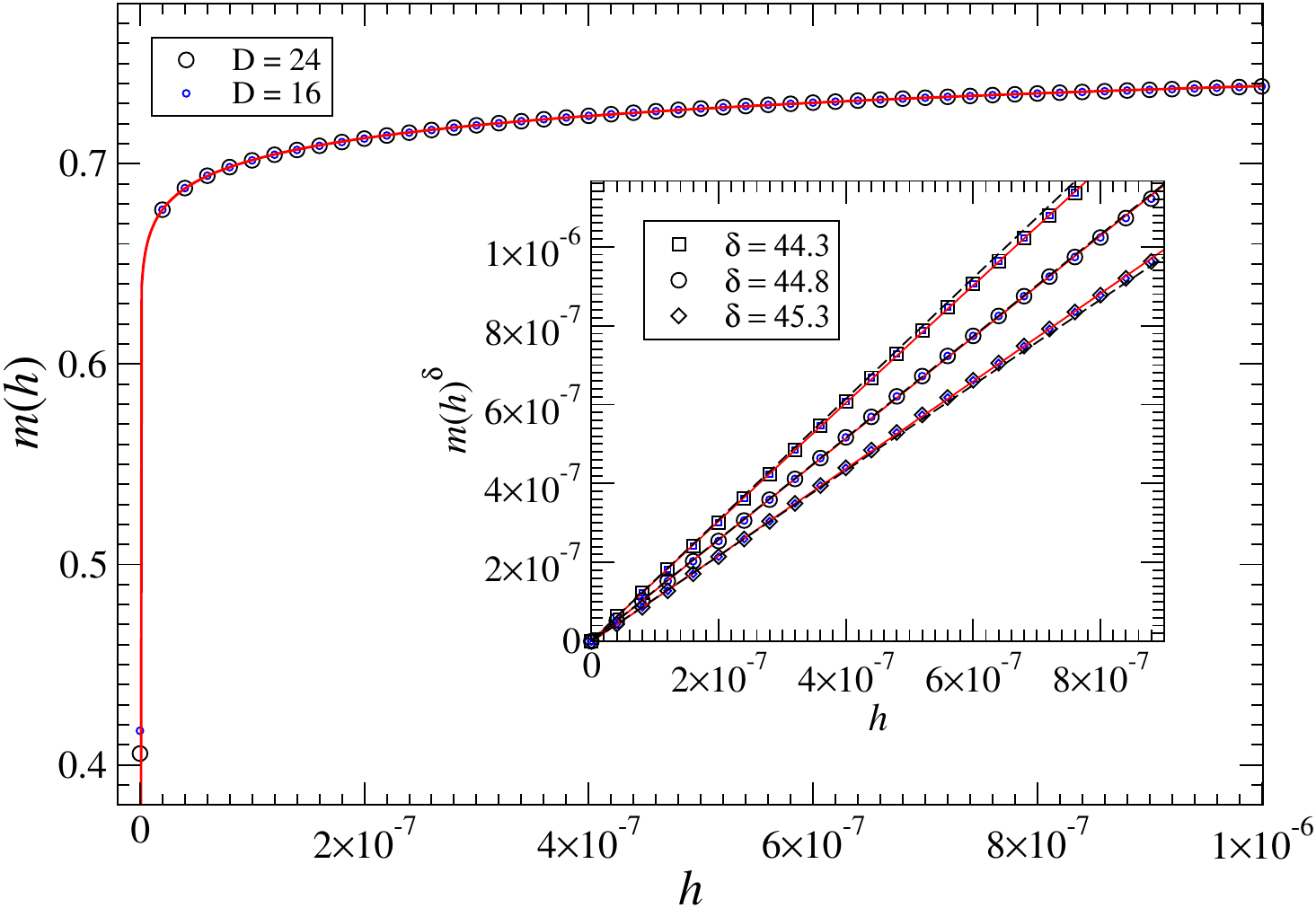}
\caption{
(Color online) 
A detailed view of the global magnetization $m(h)$ as a function of the external field $h$ at fixed temperature $T=T_{\rm c}^{~}=1.3171724$. 
The critical exponent $\delta$ was found to be $\delta \approx 44.8$. 
Inset: Critical scaling of $m(h)^{\delta}$ at $T=T_{\rm c}^{~}=1.3171724$.
}
\label{fig:delta_global}
\end{figure}

To extract the global critical exponent $\delta$, we calculate the magnetization at the critical temperature $T = T_{\rm c}^{~}=1.3171724$ as a function of the external field $h$ from the free energy per site using the automatic differentiation, $m(h, T=T_{\rm c}^{~}) = - \partial f(h, T=T_{\rm c}^{~}) / \partial h$, see Fig.~\ref{fig:delta_global}. 
The numerical analysis was performed for very small values of the external magnetic field $h$ ranging from $h = 2 \times 10^{-8}$ up to $h = 5 \times 10^{-7}$. 
The critical exponent $\delta$ was found to be $\delta \approx 44.8$ and there is practically no difference between the results for $D = 16$ and $D = 24$. 
The linear dependence of $m(h)^{\delta}$ at $T = T_{\rm c}^{~}=1.3171724$ is shown in the inset of Fig.~\ref{fig:delta_global}.

\section{Conclusions and Discussions}
In this paper, we have built upon earlier HOTRG investigations of the phase transition of the Ising model on a fractal lattice shown in Fig.~\ref{fig:Fig_1}.
Previously, the values of only two critical exponents related to magnetization ($\beta$ and $\delta$) were obtained by means of the local impurity tensors in Ref.~\cite{2dising, APS, j1j2}. 
In this study, we have extracted the remaining four critical exponents using the local impurity tensors.
%
%
Additionally, we have leveraged automatic differentiation to extract accurate estimates of four critical exponents globally by differentiating the free energy; see Table.~\ref{tab:table1}.
%
%
The automatic differentiation was also used to accurately calculate the local specific heat and magnetic susceptibility as derivatives of the local bond energy and magnetization, respectively. 
Correlations near the critical temperature were analyzed by means of two impurity tensors inserted into the system. 
From the correlations we obtained the correlation lengths from which we extracted the values of the critical exponent $\nu$ and $E = (d - 2 + \eta)$. 

\begin{table}[]
\caption{
Comparison of local and global critical exponents on a fractal lattice Ising model. 
Local and global exponents are listed in the first and second rows, respectively. 
The values of $\nu$ and $E = (d - 2 + \eta)$ (where $d$ stands for the spatial dimension) in the last two columns were extracted from the correlations obtained by the local (impurity) method only and therefore are missing in the second row. 
All listed values are obtained with the bond dimension $D=24$.
}
\begin{tabular}{llrlrlrlrlclc}
\hline
Type &
   &
  \multicolumn{1}{c}{$\alpha$} &
   &
  \multicolumn{1}{c}{$\beta$} &
   &
  \multicolumn{1}{c}{$\gamma$} &
   &
  \multicolumn{1}{c}{$\delta$} &
   &
  $\nu$ &
   &
  $(d - 2 + \eta)$ \\ \hline
Local     &  & $-0.87$ &  & $0.01383$ &  & $2.826$ &  & $204.93$ &  & \multicolumn{1}{r}{$1.60$} &  & $0.02$ \\
Global     &  & $-0.824$ &  & $0.0629$ &  & $2.764$ &  & $44.8$  &  & --                         &  & --     \\ 
\hline
\end{tabular}
\label{tab:table1}
\end{table}
The critical exponents are not entirely independent of each other~\cite{Baxter}; they are expected to satisfy the following rules given by various scaling assumptions
\begin{align}
\gamma &= \beta \left( \delta - 1 \right) \label{scaling1} \, , \\
\alpha + 2\beta + \gamma &= 2 \label{scaling2} \, , \\
\left(2 - \eta \right) \nu &= \gamma \label{scaling3} \, , \\
d \nu &= 2 - \alpha \label{hyperscaling} \, . 
\end{align}
The last equation, which involves the system (lattice) dimension $d$, can be derived by making further assumptions, known as the hyperscaling hypothesis. 
Moreover, if just two independent critical exponents are known, the remaining exponents can be derived from Eqs.~\eqref{scaling1}--\eqref{hyperscaling}, assuming a value of the spatial dimension $d$ of the system. 
The position dependence of the critical behavior on the fractal lattice was already manifested by a slight discrepancy between the single-site impurity approach used in Refs.~\cite{2dising, APS} and the multi-site impurity average approach used in Ref.~\cite{j1j2}.
In the case of single-site impurities, the observation site is kept far from the ``outer'' boundary of the system.
Note that this approach was also used in the present study when defining the local impurities.
In the case of the multi-site average employed in Ref.~\cite{j1j2}, the impurity was obtained (for convenience) as an average over four sites in the center of the 12-cluster defined by Eq.~\eqref{T_ext}. 
This partial average covers one-third of all sites; however, it does not represent the whole system, as the measurement sites are concentrated in the centers of local clusters only. 
In the case of the magnetic critical exponents, the single-site measurements yielded $\beta \approx 0.0138$ and $\delta \approx 205$, whereas partial averaging yielded $\beta \approx 0.0154$ and $\delta \approx 185$.
Notice that the partial averaging yielded $\beta$ around $10\%$ larger and $\delta$ around $10\%$ smaller than the respective exponents obtained by single-site measurements.
Comparing with Table.~\ref{tab:table1}, one can see that the partially averaged values of $\beta$ and $\delta$ lie somewhere between the local and global values, which is to be expected. 
Significant differences between the local and global exponents were observed; see Table.~\ref{tab:table1}. 
Although in the case of the exponents $\alpha$ and $\gamma$, the local and global observation yielded similar values, the differences in the case of $\beta$ and $\delta$ are large.
Large differences between the global and local critical behavior we observed (see Table~\ref{tab:table1}) show a strong position dependence on the fractal lattice. 
This should not be surprising as the fractal structure is not homogeneous, and the local behavior is similar to the surface behavior observed in Refs.~\cite{btrg, btnr}.
Let us emphasize that the fractal lattice accommodates large ``outer'' and ``inner'' boundaries at each level which inevitably leads to some degree of position dependence.
We assume that the significant difference between the local and global quantities can be explained by the contribution of surface behavior, which we intend to study quantitatively elsewhere. 
Despite the considerable differences between the local and global critical behavior, the scaling relations Eq.~\eqref{scaling1} and Eq.~\eqref{scaling2} seem to be approximately satisfied for both sets of local and global exponents. 
Given our numerical estimates of the exponents $\beta$ and $\delta$, we evaluate the relative difference between the values obtained from the scaling relations and the numerical values of the exponents $\gamma$ and $\alpha$. 
The evaluations for the local and global exponents are listed in the first and second row of Table~\ref{tab:table2}, respectively. 
For the exponent $\gamma$ we define $\gamma_{\beta\delta} = \beta \left(\delta - 1\right)$ (\textit{cf.} Eq.~\eqref{scaling1}), whereas for $\alpha$ we define $\alpha_{\beta\delta} = 2 - \beta \left(\delta + 1\right)$ (\textit{cf.} Eq.~\eqref{scaling1} and Eq.~\eqref{scaling2}). 

In the case of the local exponents, we also calculate dimensions $d_1^{~} = \gamma / \nu + E$ (\textit{cf.} Eq.~\eqref{scaling3}) and $d_2^{~} = (2 - \alpha) / \nu$ (\textit{cf.} Eq.~\eqref{hyperscaling}), which are to be compared with the Hausdorff dimension $d_{\rm H}^{~} = \log 12 / \log 4 \approx 1.79248$ in the last two columns of Table~\ref{tab:table2}. 
Especially the hyperscaling relation with the exponents $\alpha$ and $\nu$ in Eq.~\eqref{hyperscaling} is satisfied remarkably well since $d_2 \approx d_{\rm H}^{~}$.
For $d_{\rm eff}$ defined in Eq.~\eqref{d_eff} (which comes from combining Eq.~\eqref{scaling2} with Eq.~\eqref{hyperscaling}), we get $d_{\rm eff} \approx 1.782$, which yields the relative difference between $d_{\rm eff}$ and $d_{\rm H}^{~}$ to be around $0.6\%$.
\begin{table}[]
\caption{
Evaluation of the scaling relations. 
For the exponent $\gamma$ we define $\gamma_{\beta\delta} = \beta \left(\delta - 1\right)$ (first column), whereas for $\alpha$ we define $\alpha_{\beta\delta} = 2 - \beta \left(\delta + 1\right)$ (second column). 
In the case of the local exponents, we also calculate dimensions $d_1^{~} = \gamma / \nu + E$ and $d_2^{~} = (2 - \alpha) / \nu$, which are to be compared with the Hausdorff dimension $d_{\rm H}^{~} = \log 12 / \log 4 \approx 1.79248$.
%
%
%
%
%
%
%
%
%
%
%
%
%
%
%
}
\label{tab:table2}
\begin{tabular}{llccccccc}
\hline
Type &
   &
  ${|\gamma - \gamma_{\beta \delta}|} / {\gamma_{\beta \delta}}$ &
  \multicolumn{1}{l}{} &
  $-{|\alpha - \alpha_{\beta \delta}|} / {\alpha_{\beta \delta}}$ &
  \multicolumn{1}{l}{} &
  $d_1^{~}$ &
  \multicolumn{1}{l}{} &
  $d_2^{~}$ \\ \hline
Local &  & $<0.002$ &  & $<0.03$ &  & $\approx 1.784$ &  & $\approx 1.79246$ \\
Global &  & $<0.003$ &  & $<0.07$   &  & -- &  & --   \\
\hline
\end{tabular}
\end{table}
The fractal lattice studied here has the same Hausdorff dimension $d_{\rm H}^{~}$ as the Sierpinski carpet SC$(4,2)$.
When comparing the critical exponents we obtained here globally with those from the Monte Carlo studies for SC$(4,2)$ in Refs.~\cite{Carmona, Monceau, Bab2}, it seems that there are large differences.
However, the critical exponent $\alpha$ is negative in both cases. 
Also, we obtained $\beta$ close to the value reported in the newer short-time critical dynamic scaling study in Ref.~\cite{Bab2}.
In the case of fractals, the values of the critical exponents depend on the details of the lattice structure, such as lacunarity and connectivity.
It is well known that only a weak version of universality survives on self-similar structures such as fractals~\cite{complex_sc}.
However, the relations between the exponents may still be preserved.

\begin{acknowledgments}
This work was supported by the National Science and Technology Council, the Ministry of Education (Higher Education Sprout Project No.~NTU-111L104022), and the National Center for Theoretical Sciences of Taiwan. 
The support received from the project SAS-MOST Joint Research Projects 108-2112-M-002-020-MY3 is acknowledged. 
\end{acknowledgments}

\appendix


\begin{thebibliography}{99}
%
\bibitem{Domb_Green} {\it Phase Transitions and Critical Phenomena}, Vols. 1--20, edited by C.~Domb, M.~S.~Green, and J.~Lebowitz (Academic Press, New York, 1972--2001).
%
\bibitem{Baxter} R.~J.~Baxter, \textit{Exactly Solved Models in Statistical Mechanics} (Academic Press, London, 1982).
%
\bibitem{Carmona} J.~M.~Carmona, Umberto Marini Bettolo Marconi, J.~J.~Ruiz-Lorenzo, A.~Taranc\'on, Phys. Rev. B {\bf 58}, 14387 (1998).
%
\bibitem{Monceau} P.~Monceau, M.~Perreau, Phys. Rev. B {\bf 63}, 184420 (2001).
%
\bibitem{Bab1} M.~A.~Bab, G.~Fabricius, and E.~V.~Albano, Phys. Rev. E {\bf 71}, 036139 (2005).
%
\bibitem{Bab2} M.~A.~Bab, G.~Fabricius, and E.~V.~Albano, Physica A {\bf 388}, 370 (2009).
%
\bibitem{HOTRG} Z.~Y.~Xie, J.~Chen, M.~P.~Qin, J.~W.~Zhu, L.~P.~Yang, and T.~Xiang, Phys. Rev. B {\bf 86}, 045139 (2012). 
%
\bibitem{2dising} J.~Genzor, A.~Gendiar, and T.~Nishino, Phys. Rev. E {\bf 93}, 012141 (2016).
%
\bibitem{APS} J.~Genzor, A.~Gendiar, and T.~Nishino, Acta Phys. Slovaca {\bf 67}, 85 (2017).
%
\bibitem{j1j2} J.~Genzor, A.~Gendiar, Y.-J.~Kao, Phys. Rev. E {\bf 105}, 024124 (2022).
%
\bibitem{carpet} J.~Genzor, A.~Gendiar, and T.~Nishino, arXiv:1904.10645.
%
\bibitem{gasket} R.~Krcmar, J.~Genzor, Y.~Lee, H.~\v{C}en\v{c}arikov\'a, T.~Nishino, and A.~Gendiar, Phys. Rev. E {\bf 98}, 062114 (2018).
%
\bibitem{ad1} H.-J.~Liao, J.-G.~Liu, L.~Wang, and T.~Xiang, Phys. Rev. X {\bf 9}, 031041 (2019).
%
\bibitem{ad2} B.-B.~Chen, Y.~Gao, Y.-B.~Guo, Y.~Liu, H.-H.~Zhao, H.-J.~Liao, L.~Wang, T.~Xiang, W.~Li, and Z.-Y.~Xie, Phys. Rev B {\bf 101}, 220409(R) (2020).
%
\bibitem{TRG} Z.~Gu, M.~Levin and X.~Wen, Phys. Rev. B {\bf 78}, 205116 (2008).
%
\bibitem{Correlations} N.~Nakamoto and S.~Takeda, Sci. Rep. Kanazawa Univ. {\bf 60}, 11 (2016).
%
\bibitem{hosvd} L. de Lathauwer, B. de Moor, J.~Vandewalle, SIAM J. Matrix Anal. Appl. {\bf 21}, 1324 (2000).
%
\bibitem{btrg} S.~Iino, S.~Morita, N.~Kawashima, Phys. Rev. B {\bf 100}, 035449 (2019).
%
\bibitem{btnr} S.~Iino, S.~Morita, N.~Kawashima, Phys. Rev. B {\bf 101}, 155418 (2020).
%
\bibitem{complex_sc} Y.-K.~Wu and B.~Hu, Phys. Rev. A {\bf 35}, 1404 (1987).
%
\end{thebibliography}
\end{document}